# Neutron lifetime measurements with the big gravitational trap for ultracold neutrons

A. P. Serebrov,[1]* E .A. Kolomensky,[1] A. K. Fomin,[1] I. A. Krasnoschekova,[1] A. V. Vassiljev,[1] D. M. Prudnikov,[1] I. V. Shoka,[1] A. V. Chechkin,[1] M. E. Chaikovskiy,[1] V. E. Varlamov,[1] S. N. Ivanov, A. N. Pirozhkov,[1] P. Geltenbort,[2] O. Zimmer,[2] T. Jenke,[2] M. Van der Grinten,[3] M. Tucker,[3]

[1] *Petersburg Nuclear Physics Institute, NRC "Kurchatov Institute", RU-188300 Gatchina, Leningrad District, Russia*

[2] *Institut Max von Laue Paul Langevin, 71 avenue des Martyrs, F-38042 Grenoble Cedex 9, France*

[3] *Science and Technology Facilities Council, Rutherford Appleton Laboratory, Harwell Campus, Didcot, Oxon, OX11 0QX, UK*

Abstract

Neutron lifetime is one of the most important physical constants which determines parameters of the weak interaction and predictions of primordial nucleosynthesis theory. There remains the unsolved problem of a 3.9σ discrepancy between measurements of this lifetime using neutrons in beams and those with stored neutrons (UCN). In our experiment we measure the lifetime of neutrons trapped by Earth's gravity in an open-topped vessel. Two configurations of the trap geometry are used to change the mean frequency of UCN collisions with the surfaces – this is achieved by plunging an additional surface into the trap without breaking the vacuum. The trap walls are coated with a hydrogen-less fluorine-containing polymer to reduce losses of UCN. The stability of this coating to multiple thermal cycles between 80 K and 300 K was tested. At 80 K, the probability of UCN loss due to collisions with the trap walls is just 1.5% of the probability of beta-decay. The free neutron lifetime is determined by extrapolation to an infinitely large trap with zero collision frequency. The result of these measurements is $\tau_n = 881.5 \pm 0.7_{stat} \pm 0.6_{syst}\ s$ which is consistent with the conventional value of 880.2±1.0s presented by the Particle Data Group. Future prospects for this experiment are in further cooling to 10 K which will lead to an improved accuracy of measurement. In conclusion we present an analysis of currently-available data on various measurements of the neutron lifetime.

## 1. Introduction

An accurate measurement of the neutron lifetime is of great importance for the physics of elementary particles and cosmology. The decay of a free neutron into a proton, electron, and antineutrino is determined by the parameters of the weak interaction, namely, the transition of a $d$ quark into a $u$ quark. In the Standard Model of elementary particles, quark mixing is described by the Cabibbo-Kobayashi-Maskawa matrix, and unitarity must indicate the completeness of our ideas about the quark model of particles. For example, for the first matrix row:

$$|V_{ud}|^2 + |V_{us}|^2 + |V_{ub}|^2 = 1 \qquad (1)$$

Where $V_{ud}$, $V_{us}$, и $V_{ub}$ are the matrix elements related to mixing of a $u$ quark with a $d$, $s$ and $b$ quarks respectively. The values of these matrix elements are determined from the weak decay of the corresponding quarks. The theoretical description of the beta decay of a free neutron is simpler than nuclear decay, so it is preferable to use this option to determine the element $V_{ud}$. The neutron half-life $\tau_{1/2}$ is determined by the following equation [1]:

$$f\tau_{1/2}(1+\delta_R') = \frac{K}{|V_{ud}|^2 G_F^2 (1+3\lambda^2)(1+\Delta_R)} \qquad (2)$$

where $f = 1.6886$ is a phase space factor; $\delta_R' = 1.466 \cdot 10^{-2}$ is a model-independent radiative correction calculated with the precision of $9 \cdot 10^{-5}$ [2, 3]; $\Delta_R = 2.40 \cdot 10^{-2}$ is a model dependent inner radiative correction calculated with the precision $8 \cdot 10^{-4}$ [4, 5]; $\lambda = G_A/G_V$ is the ratio of the axial-vector weak coupling constant to the vector weak coupling constant and is determined from the measured angular correlations in neutron beta-decay; $G_F$ is the Fermi weak coupling constant determined from β-decay; and $K = \hbar(2\pi^3 \ln 2)(\hbar c)^6/(m_e c^2)^5$. The general formula for $|V_{ud}|^2$ as a function of neutron lifetime $\tau_n$ and $\lambda$ takes the form [1, 3]:

$$|V_{ud}|^2 = \frac{(4908.7 \pm 1.9)\ s}{\tau_n(1+3\lambda^2)} \qquad (3)$$

where the accuracy in calculation of the radiative corrections has been included. The precision of the theoretical calculation is $4 \cdot 10^{-4}$, therefore the accuracy of the neutron lifetime measurement must be better than $10^{-3}$. The precision of neutron beta-decay asymmetry measurements is inferior to neutron lifetime measurements and sets the error in the $V_{ud}$ element.

The advance in precision of measurements of neutron lifetime and the λ parameter are also important for cosmology and astrophysics since they are used in the theory of the evolution of the universe after the Big Bang and in describing the processes that determine the energy production in stars.

According to the currently accepted conceptual model, the primordial nuclear content of the Universe was formed during the Big Bang [6]. The theory of Big Bang nucleosynthesis has been developed to describe that process. It is believed that approximately 100 seconds after the origin of the universe, its temperature was $T > 10^{10} K$ $(E > 1\ MeV)$ and all leptons, hadrons, and photons were in thermodynamic equilibrium which was later violated as the temperature cooled to below $\sim 0.8\ MeV$ [7]. The relative concentration of protons and neutrons during the cooling process can be described as: $n_n/n_p \sim \exp(-\Delta m/T_f)$, where $\Delta m$ is the mass difference between the neutron and the proton, and $T_f$ is a freeze-out temperature of the weak reactions. This ratio freezes at a value of 1/5 and further change occurs only due to the beta decay of the neutron. Thus, by putting various neutron lifetimes into the model of nucleosynthesis, by the time the temperature decreases sufficiently to form heavier nuclei we obtain different ratios of protons to neutrons and therefore different concentrations of primary helium and other elements.

The abundance of helium and other elements in the Universe is determined from spectroscopic observations and from fluctuations in the cosmic microwave background (see, for example, [8, 9]). Although the changes in the concentration of elements caused by the uncertainty in the knowledge of $\tau_n$ are smaller than the errors in the observational data [10], they are already comparable, and further increase in the accuracy of the observational data will require a more accu-rate determination of the neutron lifetime. At the same time, the existing disagreement between the calculated and the observed $^4He$ abundance [11], together with the analysis of the anisotropy of the microwave background [12], can also be interpreted as the presence of additional relativistic degrees of freedom such as, for example, a sterile neutrino.

## 2. Theoretical concepts of the experiment

In our experiment the neutron lifetime is measured using stored UCN (ultracold neutrons) trapped by Earth's gravity in an open-topped vessel. The probability of loss from this trap is just 1.5% of the probability of beta-decay, so our result is close to a direct measurement of the free neutron lifetime. An additional surface, which approximately doubles collision (loss) frequency of UCN with walls, can be plunged into the trap to enable the correction due to losses of the neutrons from interaction with the walls to be taken into account.

The possibility of storing neutrons for a long time allows us to carry out measurements according to the following scheme. An evacuated, cooled, material trap is loaded with UCN, having a specially-prepared spectrum of energies. These neutrons are then held in this trap for a specific period of time. After this holding period, the trap is emptied onto a detector. The number of neutrons registered by the detector depends on the holding period, and is determined by the free neutron lifetime and losses due to collisions with the walls of the trap. Using measurements of the neutron storage time in the trap with different holding periods, and considering the theoretical dependence of the number of neutrons on time, we can calculate the lifetime of the free neutron.

Here we present the basic theoretical model of the processes. For simplicity, the model includes some assumptions concerning the interaction of UCN with walls. In the experiment, the actual conditions differ from the ideal model. These differences cause systematic errors in our experiment and will be discussed below. The assumptions in our model are:

1. Neutrons move freely between collisions affected only by gravity, we assume that there are no collisions with residual gas nuclei.
2. The energies of the neutrons do not change due to collisions with walls.
3. Losses of neutrons due to collisions with walls are described by the model of interaction with potential barrier.
4. The coating of the walls is homogeneous.

At first we consider set of mono-energetic neutrons. Using this assumption the number of neutrons remaining in the trap as a function of time is described by the expression:

$$N(t,E) = N_0 \exp(-t/\tau_{st}(E)) \quad (4)$$

where $N_0$ is the amount of trapped neutrons at the beginning of holding period, $N(t)$ is the number of neutrons after time $t$, and and $\tau_{st}$ is the lifetime of neutrons stored in the trap (storage time), defined by the expression:

$$\tau_{\mathrm{st}}^{-1}(E) = \tau_{\mathrm{n}}^{-1} + \tau_{loss}^{-1}(E) \quad (5)$$

where $\tau_{\mathrm{n}}^{-1}$ is the rate of β-decay, and $\tau_{loss}^{-1}$ is the rate of loss due to collisions with the walls of the trap.

By considering two holding periods, we can obtain the stored neutron lifetime without needing to know the initial number of neutrons:

$$\tau_{st} = (t_2 - t_1)/\ln(N_1/N_2) \quad (6)$$

where $N_1$ and $N_2$ are the numbers of neutrons at moments $t_1$ and $t_2$ correspondingly.

In the model, neutron loss due to collisions with the walls arises from the imaginary part of the potential barrier. To calculate loss probability (without gravity), we use the expression:

$$\tau_{loss}^{-1} = \mu(T,E)\nu(E) \quad (7)$$

where $\mu(T,E)$ is the probability that a UCN is lost at each collision – which depends on UCN energy and temperature of the wall, and $\nu(E)$ is the frequency of UCN collisions with the walls. Without the gravitational potential, the neutron kinetic energies do not depend on height.

The UCN collision loss function $\mu(T,E)$, assuming that the potential is a square barrier with real $U_0$ and imaginary $W$ parts, can be represented in the following well-known form [13]:

$$\mu(y) = \frac{2\eta}{y^2}\cdot\left(\arcsin y - y\sqrt{1-y^2}\right) \approx \begin{cases}\pi\eta, & y\to 1\\ \frac{4}{3}\eta y, & y \ll 1\end{cases} \quad (8)$$

where $\eta$ is the loss coefficient which is the ratio of imaginary to real parts of the potential (or scattering amplitudes) $\eta = W/U_0 = b'/b$, and $y = \sqrt{E/U_0} = v/v_{lim}$ is the ratio of neutron velocity to the maximum neutron velocity $v_{lim}$ at which full reflection from the potential barrier is still possible.

The UCN energy loss function in equation (8) is averaged over directions. Using the optical theorem, the imaginary part of the scattering amplitude can be represented as:

$$b' = \frac{\sigma_{abs} + \sigma_{upscat}(T)}{2\lambda} \quad (9)$$

The expression for the normal component $v_\perp$ of neutron velocity has a simple form:

$$\mu(v_\perp) = \frac{2\eta x}{\sqrt{1-x^2}} \quad (10)$$

where $x = v_\perp/v_{lim}$.

Capture and inelastic scattering cross-sections are proportional to neutron wavelength $\lambda$, hence $b'$ and $\eta$ do not depend on neutron energy, and are functions of wall temperature only $\eta = \eta(T)$. Therefore equation (7) takes the form:

$$\tau_{loss}^{-1} = \eta(T)\gamma(E) \quad (11)$$

where $\eta(T)$is the loss coefficient which doesn't depend on UCN energy, and $\gamma(E)$ is the effective collision frequency which depends on UCN energy and trap size. In this form, only one UCN energy-dependent parameter is required to calculate the free neutron lifetime from experimental data. It appears that inverse storage time is a linear function of $\gamma(E)$ with slope coefficient $\eta$ and a value equal to the inverse lifetime of the free neutron at $\gamma(E) = 0$. Therefore, neutron lifetime can be obtained by linear extrapolation of $\tau_{st}^{-1}$ to zero $\gamma(E)$, i.e. no collisions of the UCNs with the inner walls of the trap.

Using the linear dependence on $\gamma(E)$ we can obtain neutron lifetime using measurements of neutron storage times for two different $\gamma(E)$.

$$\tau_1^{-1} = \tau_n^{-1} + \eta\gamma_1 \ ; \ \tau_2^{-1} = \tau_n^{-1} + \eta\gamma_2 \quad (12)$$

So:

$$\eta = (\tau_2^{-1} - \tau_1^{-1})/(\gamma_2 - \gamma_1) \ ; \quad \tau_n^{-1} = [(\tau_1^{-1} + \tau_2^{-1}) - \eta(\gamma_1 + \gamma_2)]/2 \quad (13)$$

where $\gamma_1$ and $\gamma_2$ are effective collision frequencies, and $\eta$ — loss coefficient. Eliminating $\eta$, we obtain:

$$\tau_n^{-1} = \tau_1^{-1} - (\tau_2^{-1} - \tau_1^{-1})/[\gamma_2(E)/\gamma_1(E) - 1] \quad (14)$$

This form exhibits the important feature of this experiment — one of the storage times is the basis for calculations of neutron lifetime, with a correction determined by the difference of storage times and a ratio and effective frequencies. This reveals why it is so important to obtain the best possible storage time.

Equation (14) represents the main idea of our experiment — to calculate the free neutron lifetime one needs to measure storage times for at least two values of $\gamma(E)$. This function $\gamma(E)$ is determined by parameters of the experimental apparatus and the physical model of the interaction. It depends on trap geometry and UCN energy. Hence there are two ways to measure storage times with two different $\gamma(E)$: using different energies or different geometric configurations. “Energy extrapolation” is where we use two different spectra and the same geometry, and "geometry extrapolation" is where we modify the trap geometry to change the effective collision frequency. In our experiment we use an additional surface, the insert, to change the total surface area. The method of geometry extrapolation has significant advantages compared to that of energy extrapolation and we discuss these further in detail.

It is important that equation (14) includes only $\gamma_2(E)/\gamma_1(E)$, since it reduces the influence of function $\mu(E)$ on the results. In the geometry extrapolation method, and ignoring the effects of gravity, the ratio of effective frequencies depends only on the configuration of the trap and insert:

$$\frac{\gamma_2}{\gamma_1} = \frac{S_2}{V_2} \Big/ \frac{S_1}{V_1} \qquad (15)$$

where $S$ is the total area of surface and $V$ is the volume of the trap. Any dependence on UCN energies is completely excluded. However, in a gravitational field, the kinetic energies and number density of the UCN depend on height. The loss rate can still be calculated using equation (7). Collision frequency at the unit of area $dS$ equals to the UCN flux towards the trap walls: $1/4\,v\rho(v)dS$, where $\rho(v)$ is the UCN number density which is a function of neutron velocity $v$. In a gravitational field, the UCN number density is proportional to $\sqrt{(E-mgh)/E}$, where $E$ is the UCN energy at the bottom of the trap, and $h$ is the height from the bottom of the trap. Since UCN energy depends on $h$, to obtain the loss rate one must integrate equation (7) over $h$ and normalize to the total amount of neutrons:

$$\tau_{loss}^{-1}(E) = \frac{\int_0^{h(E)} \mu(E-h')v(E-h')\rho(E-h')dS(h)}{4\int_0^{h(E)} \rho(E-h')dV(h)} = \eta \cdot \gamma(E) \qquad (16)$$

where $h' = mgh$ and $h(E)$ is the maximum height for energy E.

Complete exclusion of the energy dependence does not occur because the integral form of loss probability depends on $\mu(E)$. To decrease energy dependence in the geometry extrapolation, the insert has a special shape. In fact, the insert has a shape similar to the trap but with smaller size. The perfect insert would increase the surface area by a uniform factor for all values of $h$. Such an effect cannot be achieved, but the shape used aims to do this for most values of $h$.

The simple form of the equations obtained above is the result of using an ideal model of the experiment. In the real experiment we must take into account various deviations from the assumptions made in the beginning of this section. We can deal with some of the problems using compensations or considering them in a MC model of the experiment. Others will form systematic errors and be discussed in section 9.

We have presented the physical concepts of our experiment. For further discussion we now introduce the geometrical features of our experimental apparatus and calculate real values of parameters $\gamma(E)$ for our particular trap.

## 3. Experimental apparatus

The measurements were carried out at the ILL research reactor at the PF2 MAM UCN beam. Our experiment was installed on a specially-prepared platform in the experimental hall [14]. Schematics of our experimental setup are shown in Fig. 1 and Fig. 2.

To carry out measurements of neutron lifetime one needs to provide special properties of the apparatus, such as high vacuum in the trap volume and a low loss coefficient for UCN collisions with walls. In order to obtain low neutron losses on collisions, the walls of the trap were covered with a hydrogen-less fluorinated polymer (Fomblin-grease UT-18 Solvay company), which has low neutron capture cross-section. UCN losses from inelastic scattering on Fomblin are strongly suppressed at low temperatures 80-100 K.

The apparatus consists of two manufactured nitrogen tanks (Fig. 2). The outer tank contains the insulating vacuum. The inner tank contains vessels for liquid nitrogen and the primary experimental components: the neutron trap and the insert. When the vessels are filled with liquid nitrogen, the walls of the inner tank act as the thermal screen. Flexible pipelines guide vaporized gaseous nitrogen toward tubes attached to the trap and insert. This keeps the temperature of the trap and insert at 80 K with 1 K accuracy, monitored by 4 thermocouples — two for each surface. The inner and outer volumes are pumped out independently using turbomolecular pumps. Residual gas pressure in the trap volume does not exceed $2 \cdot 10^{-6}$ Torr.

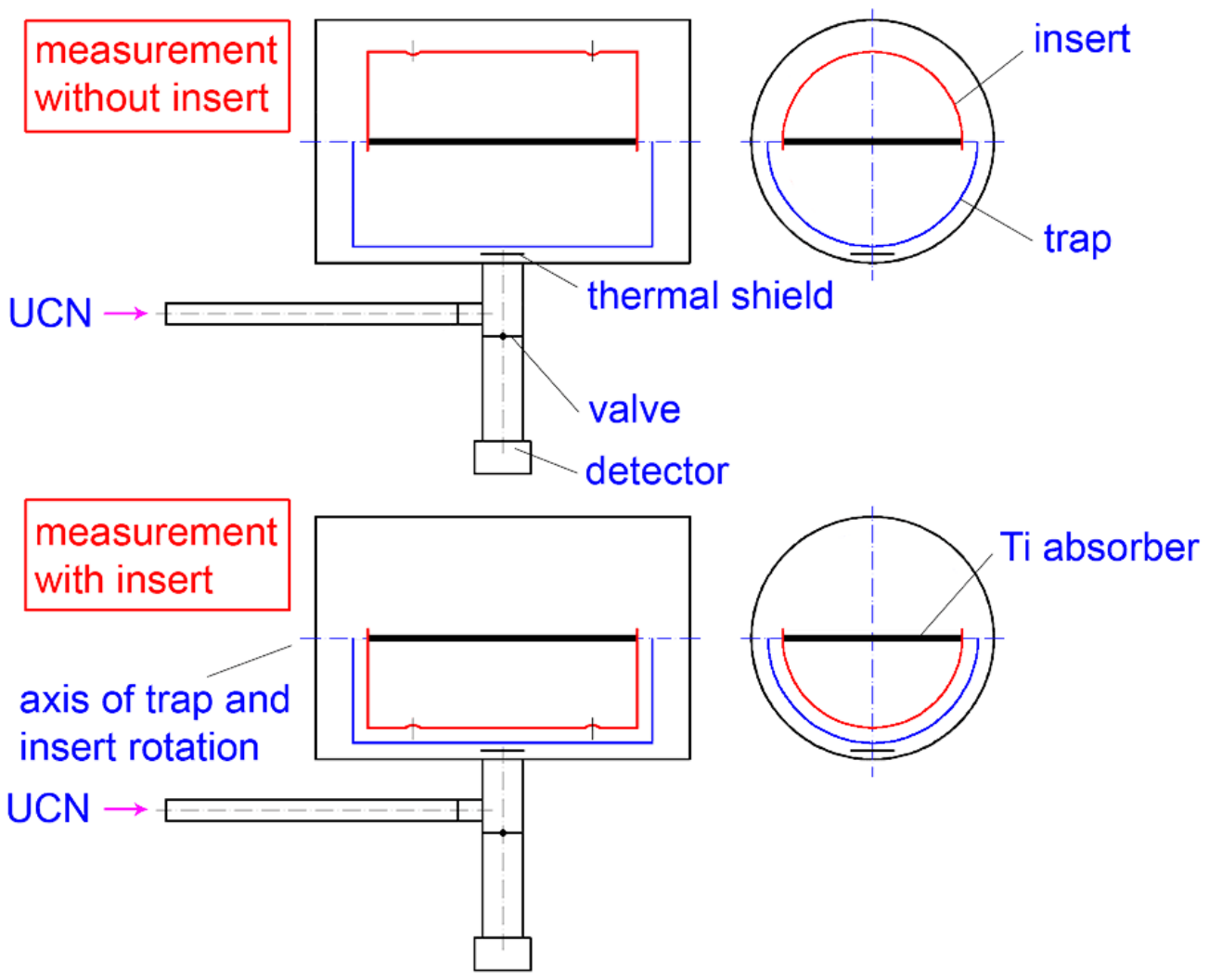

FIG. 1. Basic scheme of inner part of the apparatus.

The neutron trap is a copper half-cylinder with radius 0.7 m and length 2 m. The large size of the trap is unique, and enables the capture of numerous neutrons, greatly increasing the statistical accuracy of every measurement. For example, in our previous experiment the trap had less than one fifth of this volume [15, 16]. In order to vary neutron collision frequency, a special surface plunges into the trap. This surface (the insert) is a copper half cylinder without side walls with radius 0.6 m and length 1.8 m. The insert has holes to enable the free exchange of neutrons between trap and insert vol-umes. Trap and insert turn independently using the joint "pipe-within-pipe" shaft and stepper motors. Fig. 2 shows the general arrangement of the whole experimental apparatus and illustrates the positions of the functional parts.

Between reactor cycles 179 and 180 we added the titanium absorber. The details of its construction and influence on the experiment is discussed in section 7. The scheme of the inner part of the apparatus (Fig. 1) includes the absorber but, at first, measurements were made without it.

The trap is filled via a neutron guide, equipped with two shutters. The first one, turbine shutter (11), is closed during the measurement process to prevent neutrons from the turbine reaching the detector. The second, shutter (13), passes neutrons towards the trap during the filling. The detector shutter (12) is open during the measurement phases to pass neutrons towards the detector.

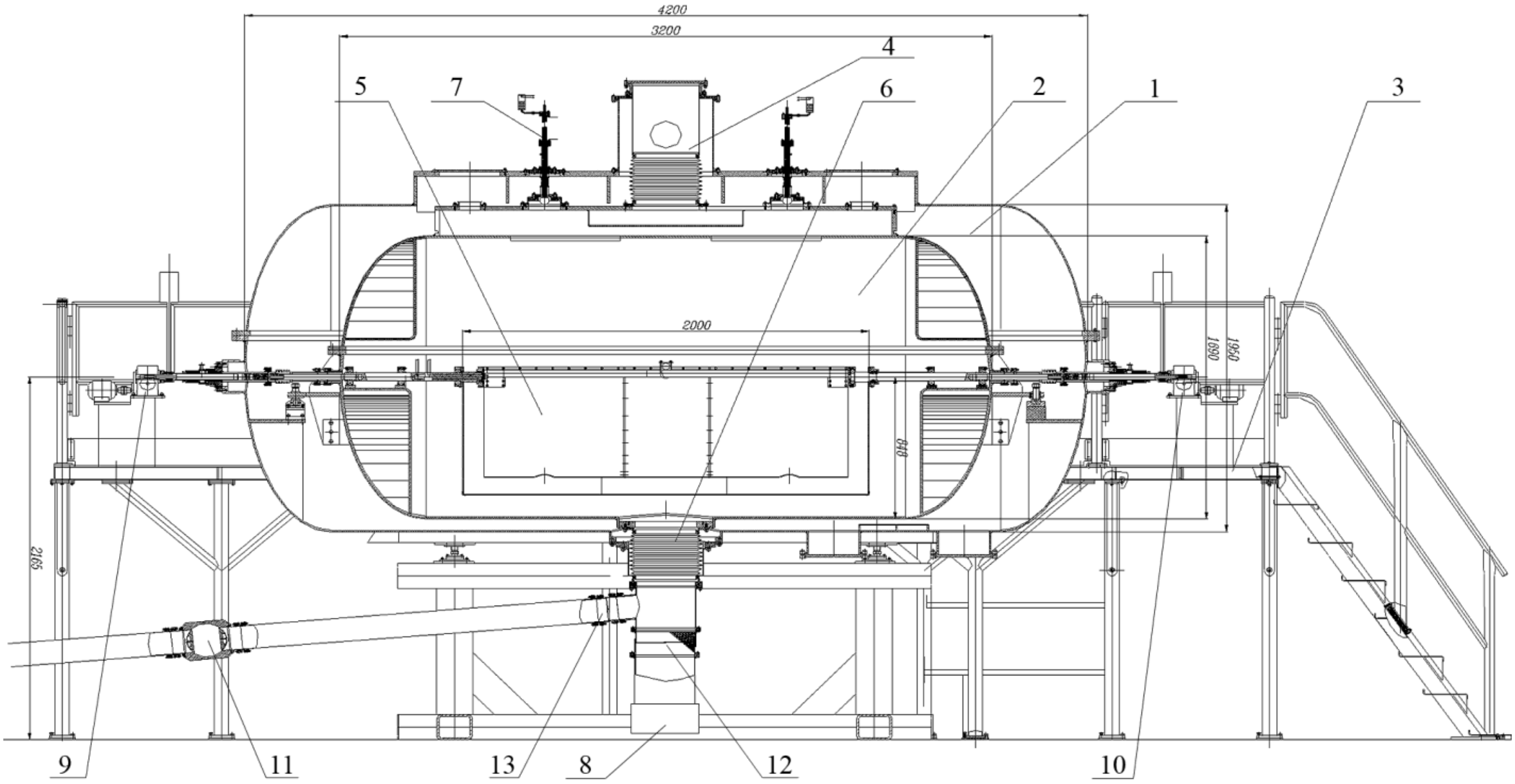

FIG. 2. 1 — external vacuum vessel; 2 — internal vacuum vessel; 3 — platform for service; 4 — gear for pumping out internal vessel; 5 — trap with insert in low position; 6 — neutron guide system; 7 — system of coating of trap and insert; 8 — detector; 9 — mechanism for turning trap; 10 — mechanism for turning insert, 11 — turbine shutter, 12 — detector shutter, 13 — neutron guide shutter.

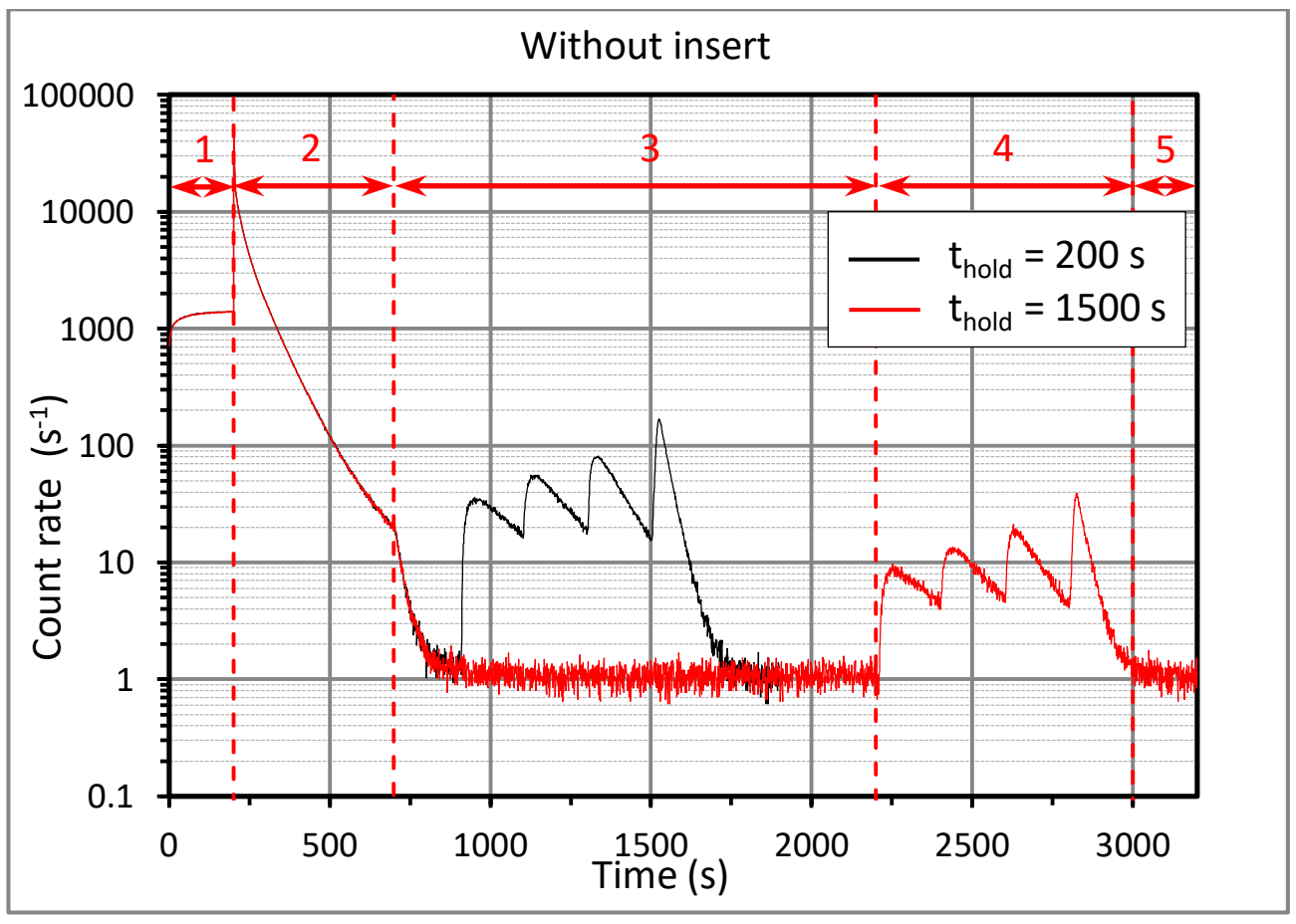

FIG. 3. Typical storage cycles for two different holding times in a trap: 1 — filling, 200 s [with time of trap rotation (50 s) to monitoring position included]; 2 — spectrum preparation, 500 s; 3 — holding, 200 s or 1500 s [with time of trap rotation (10 s) to holding position included]; 4 — emptying, with four periods of 200, 200, 200, 300 s [with time of trap rotation (10, 6, 6, 25 s) to each position included]; and 5 — measurement of the background, 200 s. Time intervals are shown for holding time 1500 s.

A typical measurement cycle is now described. At first, the trap turns to 90 degrees to be filled with UCN from the neutron guide. In this period the whole inner tank is filling with UCN. The filling process is monitored by UCN leakage through gaps in the detector shutter, which is located below the trap. After 150 seconds, when the count rate reaches a plateau (Fig. 3), neutrons are captured by turning the trap to the position with 15 degrees tilt. This rotation takes 50 seconds, after which time both inlet shutters close and the detector shutter opens (12). In order to exclude above-barrier neutrons, the trap is maintained in this position for the next 500 seconds. Following this spectrum preparation, the trap is turned to the horizontal position for the storage, which lasts for 200 or 1500 seconds. Then, the remaining UCNs are detected by turning the trap to the decanting positions. In the description of the process presented in Fig. 3 the tilt angles for decanting are 19, 24, 33 and 90 degrees. Later, a cycle with only two decanting angles was chosen as more suitable for this experiment – 24 and 90 degrees. The final 200 seconds of each cycle are used to measure the background.

The apparatus and measurement process, as described, allow us to carry out measurements of neutron lifetime in the trap for various holding periods and collision frequencies. The free neutron lifetime is obtained by linearly extrapolating to zero collision frequency of the UCNs with the trap walls.

## 4. Monte-Carlo simulation and calculation of parameter $\gamma(E)$

In Fig. 4 the dependence of the effective collision frequency $\gamma(E)$ on energy for monoenergetic UCN is presented.

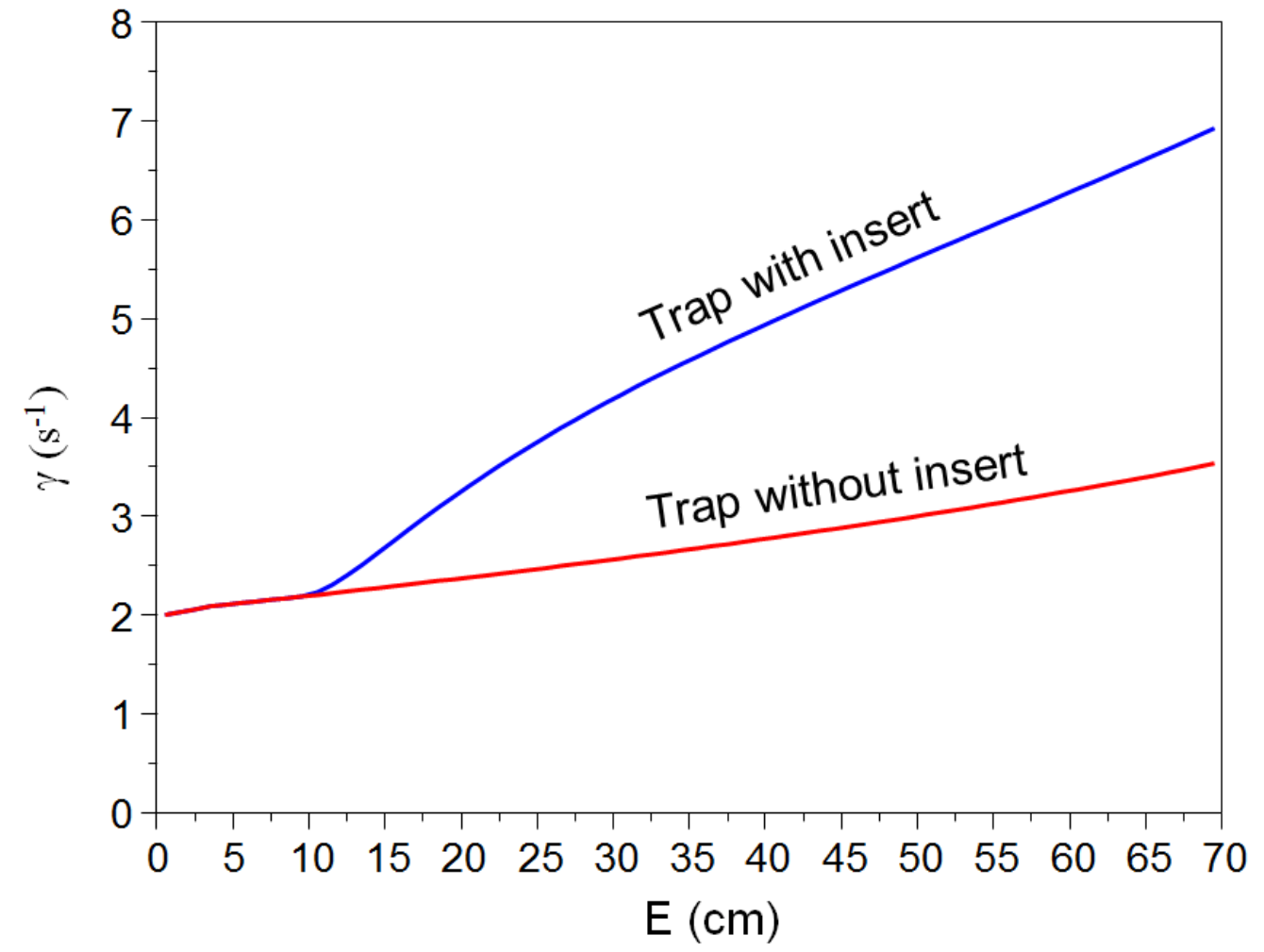

FIG. 4. The effective collision frequency $\gamma(E)$ as a function of UCN energy expressed in terms of height in the gravitational field.

It was calculated using formula (16) which considers the geometry of the trap and the gravitational potential. The lower line represents the function $\gamma(E)$ for the trap without the insert. This $\gamma(E)$ value increases slowly with energy and hence the mean collision frequency for different parts of the UCN spectrum differs only very slightly. For linear extrapolation the closer the points are, the larger the error, and this fact motivated us to use geometry extrapolation. Notice that in our model $\gamma(E)$ does not depend on $\eta$, the loss coefficient, because $\eta$ does not depend on height and can be excluded from the integral. The upper curve is the function $\gamma(E)$ for the trap with insert. This intersects the lower curve at 10 cm because the insert radius is 10 cm smaller than the trap radius and UCNs that are not energetic enough to rise 10 cm in the gravitational field do not reach the insert.

The MC simulation includes many neutron paths from the inlet neutron guide to the detector. This simulation requires parameters of the measurement cycle such as: tilt angles, insert position, time intervals, and also parameters of physical processes: the initial spectrum of UCN energies and the form of the loss function [17]. The initial spectrum is assumed to be Maxwellian and the loss probability function is determined by equation (8). The criterion for verification of the model is the equality of measured and simulated count rates. The results are shown in Fig. 5.

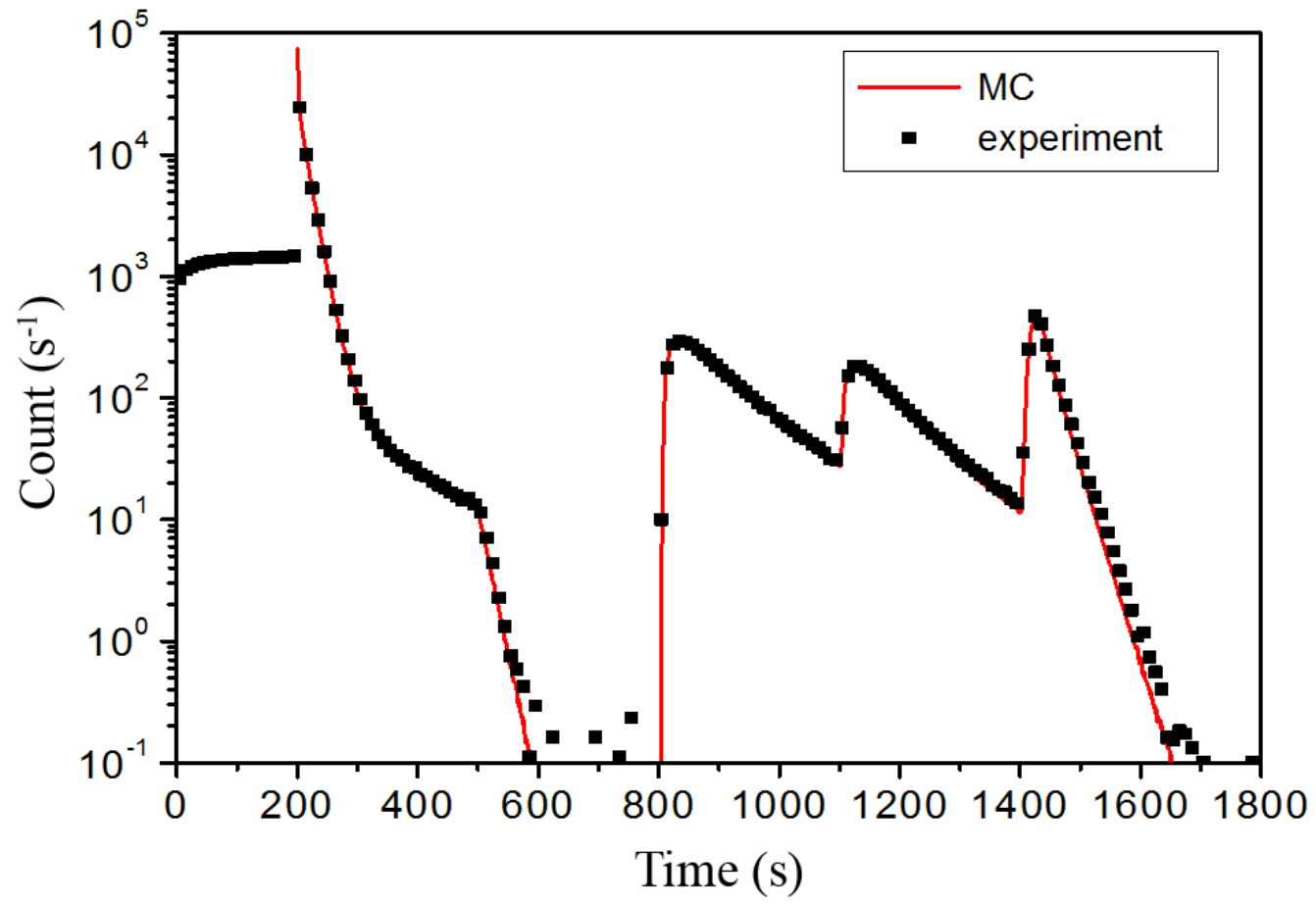

FIG. 5. Simulated and measured UCN registered by the detector.

When the modelled and experimental curves coincide, we can reconstruct the UCN energy spectrum for every cycle with high accuracy. This is the reason why we use computer modelling to obtain the energy spectrum. The spectrum of UCNs reaching the detector in the computer model is shown in Fig. 6.

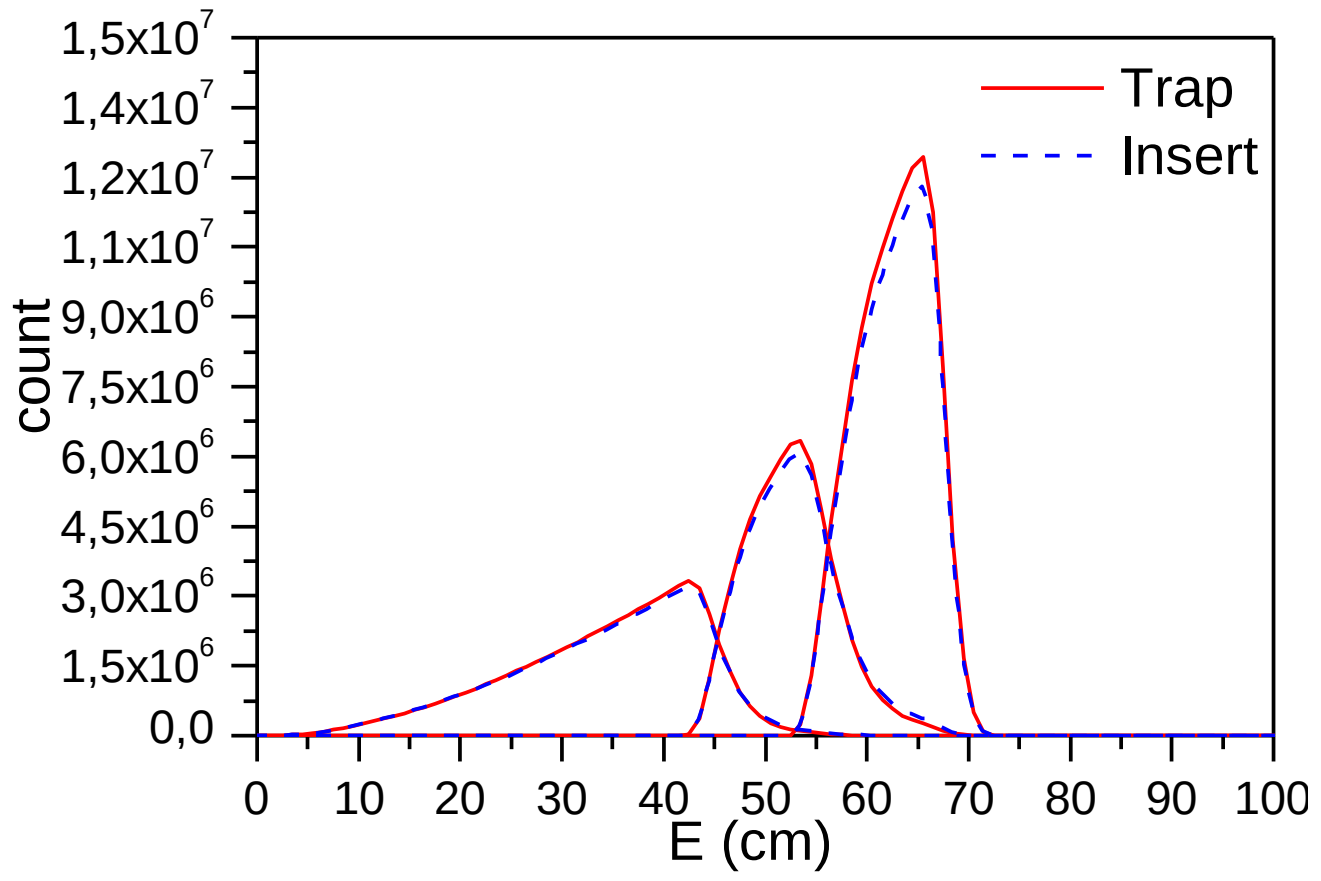

FIG. 6. Spectra of UCNs registered by the detector in the simulation when decanting from the trap at three successive angles of tilt for trap only (red line) or trap with insert (blue dashed line).

Equation (16) expresses the loss rate as a function of neutron energy. The real experiment contains a spectrum of UCN energies and gives a mean storage time. The mean effective collision frequency $\gamma_m$ is obtained from $\gamma(E)$ weighted by the spectrum of neutrons as registered by the detector and normalized to the total number of neutrons. Using the derived energy spectrum we can calculate the mean effective collision frequency $\gamma_m$ for all decanting phases in each geometric configuration. For example, with three successive decanting phases as shown in Fig. 6 we gain three points $\gamma_1^{out}$, $\gamma_2^{out}$, $\gamma_3^{out}$ for measurements with the trap only, and three points $\gamma_1^{in}$, $\gamma_2^{in}$, $\gamma_3^{in}$ for measurements with the trap and insert.

In addition to spectrum reconstruction, the computer model of the experiment provides an opportunity to make a complete simulation of the measurement process. Hence, we can predict the influences of various uncertainties in the model or measurement process on the result of measurements. To simulate the neutron lifetime, one must add another parameter in the model — the value of the free neutron lifetime. The calculations prove that points $\gamma_m$ do not depend on the free neutron lifetime over a wide range of values.

In order to check the self-consistency of the calculations the whole experiment is simulated with the obtained parameters $\gamma_m$. In the simulation we estimate the difference between the measured lifetime and that fed into the model as an input parameter, hence we represent results in the form of differences between inverse storage times and the inverse lifetime parameter. The simulated storage times are obtained using equation (6) where neutron counts are the total numbers of neutrons reaching the detector during the emptying process. There is complete agreement between the lifetime obtained by extrapolation and the value used as an input parameter.

In Fig. 7 the results of long term MC simulations with four decanting phases are shown. This was performed in order to check the accuracy of the reconstruction of the lifetime using methods of geometry and energy extrapolation. In this simulation, the loss function $\mu(E,T)$ used to calculate proba-bility of loss at collisions with walls is the same as in calculations of $\gamma_1^{out}, \gamma_2^{out}, \gamma_3^{out}, \gamma_4^{out}$ for the trap only and $\gamma_1^{in}, \gamma_2^{in}, \gamma_3^{in}, \gamma_4^{in}$ for the trap and insert. Therefore the input neutron lifetime and the lifetime obtained from the extrapolation of storage times, calculated using simulated count rates and equation (6), should be equal within the statistical accuracy of the simulation. Indeed, from the table in Fig. 7 one can conclude that the results lie within statistical error. For energy extrapolation with the insert, the accuracy is 0.33 s; and without the insert it is 0.53 s. For the joint geometry extrapolation it is 0.10 s. Thus, the accuracy of simulation is estimated to be 0.10 s. This accuracy characterizes our method of reconstructing the neutron lifetime using count rates in the detector.

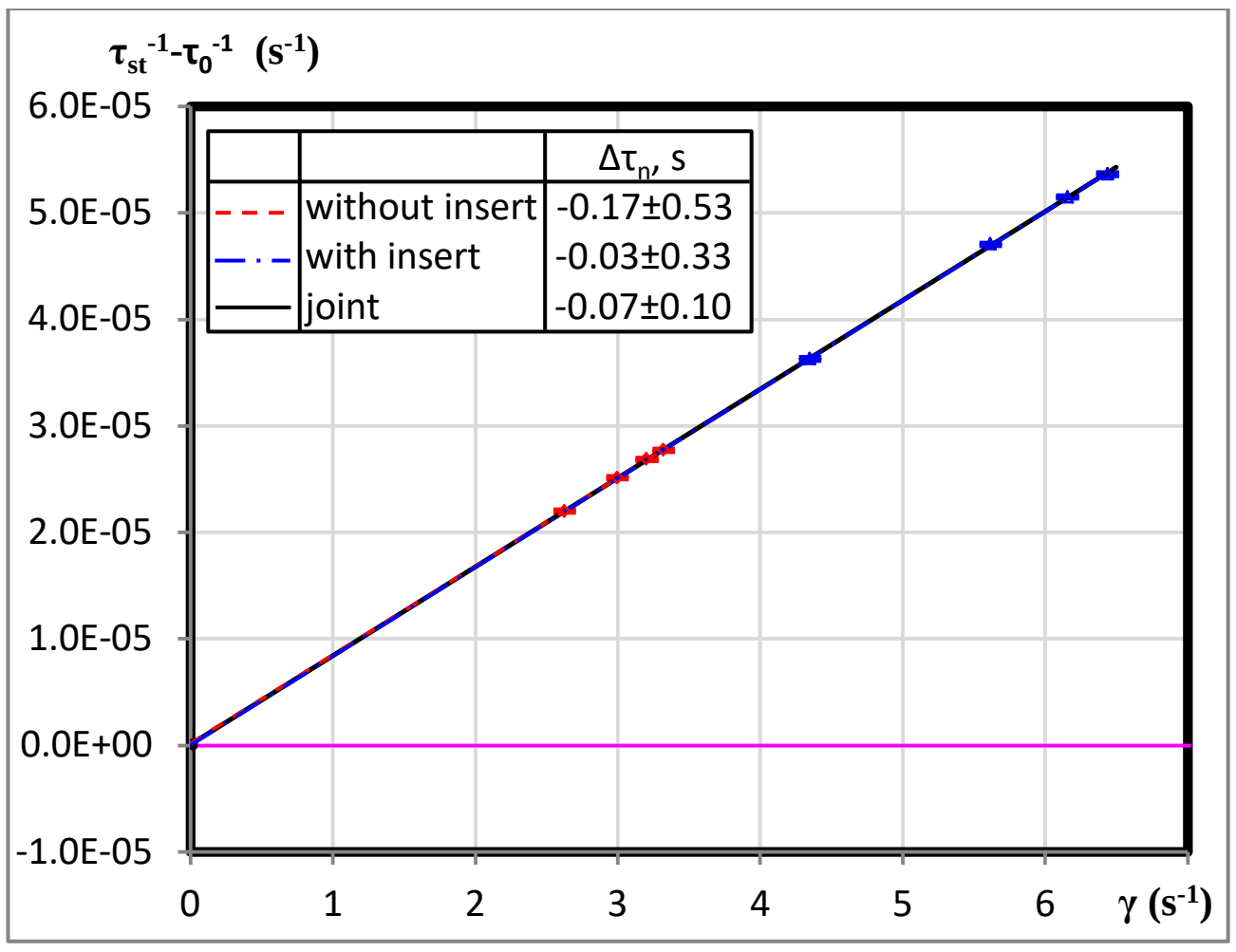

FIG. 7 The results of MC model self-consistence test.

The ideal model of the experiment was described in section 2. This model is only an approximation to reality so the accuracy of this approximation is now estimated.

To study the influence of the form of $\mu(y)$ on the results, was made the series of simulations with various loss probabilities that depended on neutron velocity, while for $\gamma$ calculations the classical function (8) was used. Simulations were made with various functions — constant, linear and quadratic functions of the parameter $y$ (Fig. 8). The results are then compared with the lifetime parameter in the model. Fig. 9 illustrates the influence of the uncertainty in the model of physical interaction with walls.

These simulations reveal that even with least probable forms of $\mu(y)$ the geometry extrapolation gives acceptable results. For example, if $\mu(y) = const$ the systematic error in energy extrapolation is -40 s, however it decreases to 1.9 s if we apply geometry extrapolation. Here we consider the discrepancy between the extrapolation result and the input neutron lifetime as the systematic error. In the simulation with quadratic $\mu(y)$, the energy extrapolation systematic error changes sign and becomes close to +20 s, while it decreases to -1.9 s when applying the geometry extrapolation. For linear $\mu(y)$ the systematic error after geometry extrapolation is only 0.2 s.

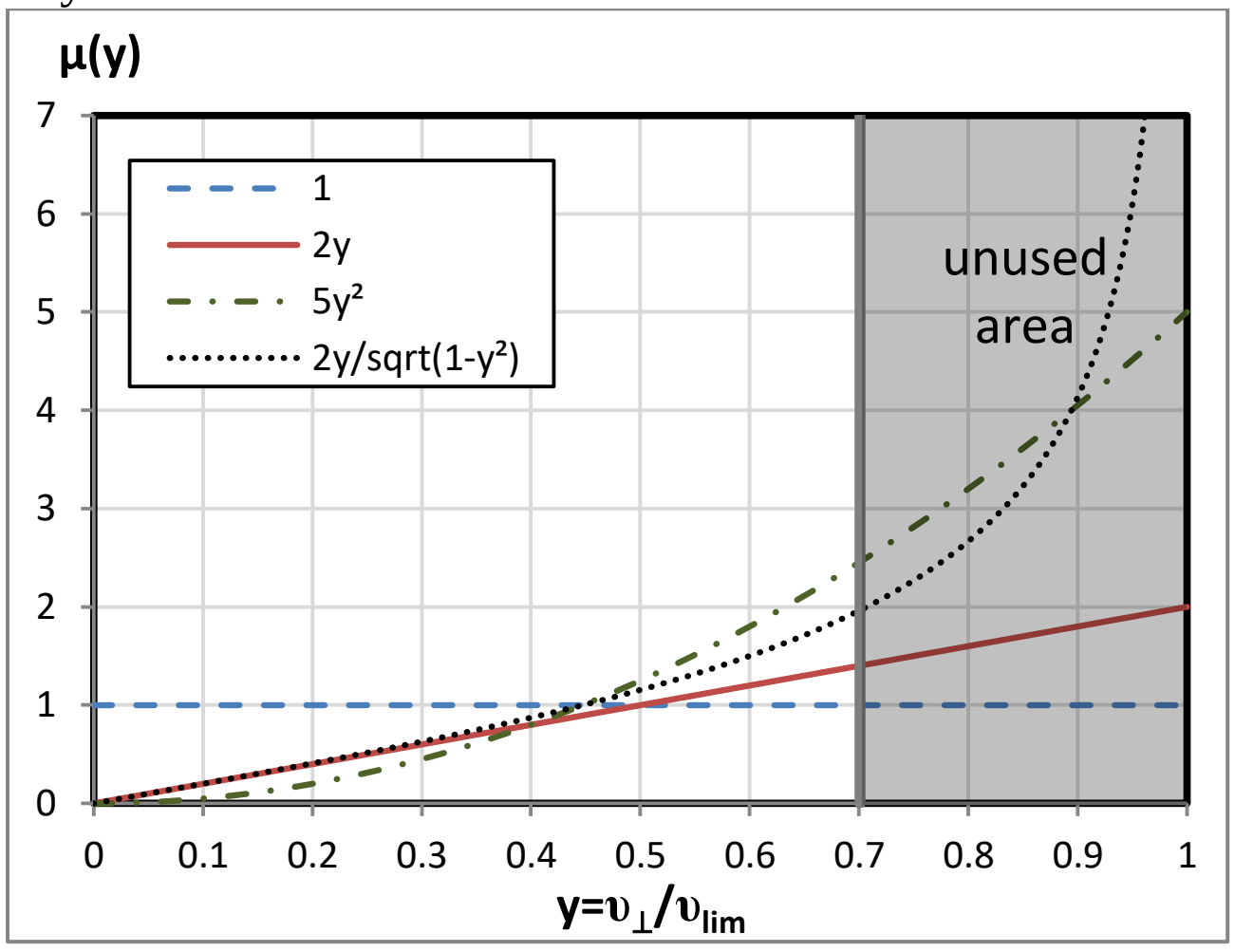

FIG. 8. Modeled μ(y) functions.

These calculations show that changes in the loss functions affect the results of geometry extrapolation significantly less than results of energy extrapolation. This is the reason we use geometry extrapolation to calculate neutron lifetime, the geometry extrapolation is much more stable.

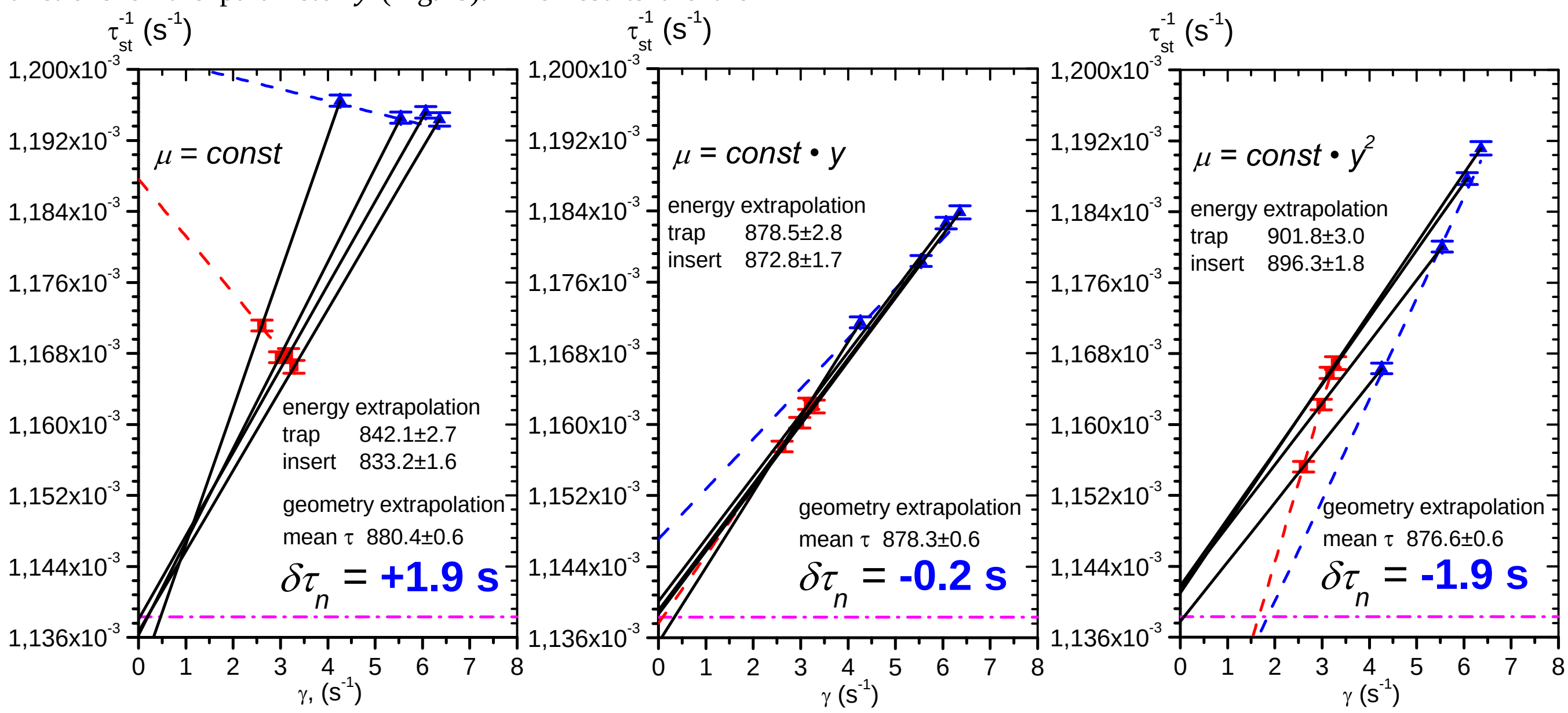

FIG. 9. The results of simulations with various forms of loss probability dependence on velocity, where $\tau_n$ = 878.5 s is the input parameter.

The real loss function might be much more complicated but the model of square barrier potential and the loss function (8) provides very good agreement of simulated count rates with measured count rates. So we believe that, using the linear approximation in the range of UCN energies trapped by this experiment, the real function is close to the model. For this reason we use the result of the simulation with the linear function to estimate systematic errors.

The agreement between the input parameter value of the neutron lifetime and the resulting neutron lifetime obtained by the extrapolation is the main test to check the computer simulation. This simulation allows us not only to obtain the results, but also to reveal and estimate the sources of systematic errors. The table of all systematic errors is presented below in section 9 subsection g.

## 5. Low temperature Fomblin

In our experiment the trap and insert are copper. Copper has high thermal conductivity and a sufficiently high Fermi potential to trap UCN with boundary velocities less than $v_{lim}$~5.7 $m/s$ But the UCN capture probability is quite high, so the copper surface was coated with Fomblin UT-18 grease (Solvay company product). This substance consists of molecules comprised of carbon and fluorine, so it has a small capture cross-section. Moreover UCN losses are strongly suppressed at low temperatures (80-100K). For the experimental method we use, it is essential to confirm that the coating of Fomblin grease remains stable during freezing and heating.

The Fomblin layer is, on average $5 \div 10\mu m$ thick which is about two orders of magnitude more than the mean length of UCN propagation in matter. For this reason we assume that the UCN loss probability on contact with Fomblin does not depend on the underlying material of the trap. In the experiment, areas with a much thinner layer or even with no covering can appear at the microscopic level. Here we call it uncoated area and estimate its influence on the result.

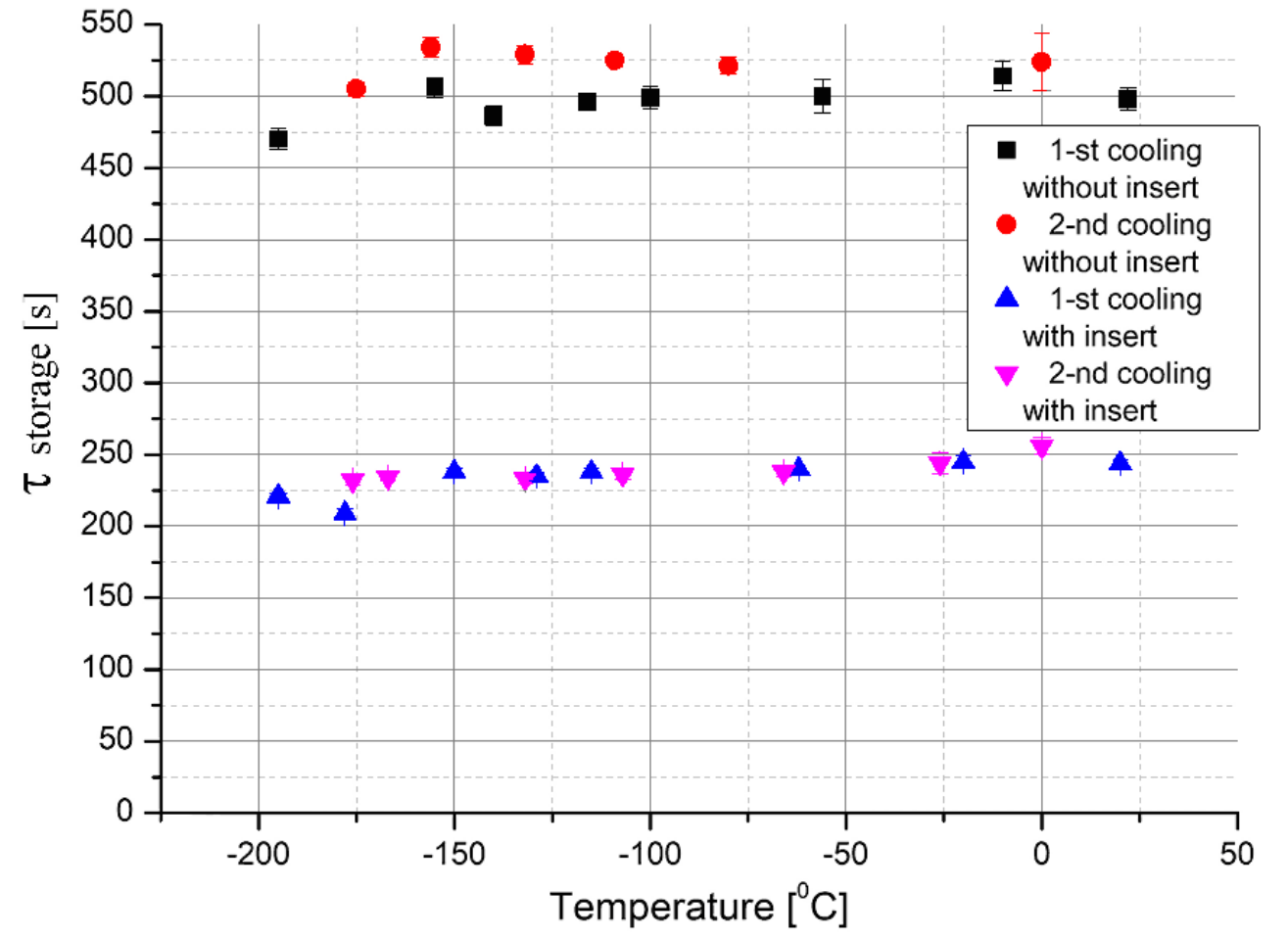

FIG. 10. Temperature dependence of the UCN storage time in the experiment with the titanium trap and insert.

Preliminary measurements of the storage time in a titanium trap and insert, coated with Fomblin grease, were carried out. Titanium practically doesn't reflect UCN due to its negative scattering length. Hence during freezing and heating even small changes in the coating significantly affect the measurements. The results are shown in Fig. 10.

One can conclude from these measurements that:

i. The coating is stable – repeating cooling and heating does not damage it in any noticeable way. It means that any uncoated area does not grow after cooling.
ii. We assume that any uncoated area in the titanium trap has at least 50% UCN capture probability and conclude that it does not exceed 0.1% of the total trap area ($\eta_{Ti+F} = 5 \cdot 10^{-4}$).

Both results are extremely important for our current experiment. The stability of the coating allows us to combine the results obtained in different reactor cycles in order to increase statistical accuracy. The estimations of uncoated area provide information to calculate systematic errors which might appear because of some inequality in trap and insert coatings.

## 6. Storage time and loss dependence on temperature

To control coating stability, we carry out measurements of storage time during every freezing and heating of the trap. Copper has a small, in comparison with titanium, loss probability for UCN collisions with walls ($\eta_{Cu} \approx 10^{-4}$). From the experiment with the titanium trap we conclude that uncoated areas of the trap and insert do not exceed 0.1% of the total and hence, for the copper trap, the uncoated area contribution to the total loss probability is $1 \cdot 10^{-7}$. UCN loss probability on Fomblin is smaller than $10^{-5}$ and temperature dependent. Using the titanium trap we could not observe that dependence because losses on the uncoated area made a much bigger contribution to the total loss probability (Fig. 10). The copper trap made observable this effect of the loss probability decreasing with temperature (Fig. 11).

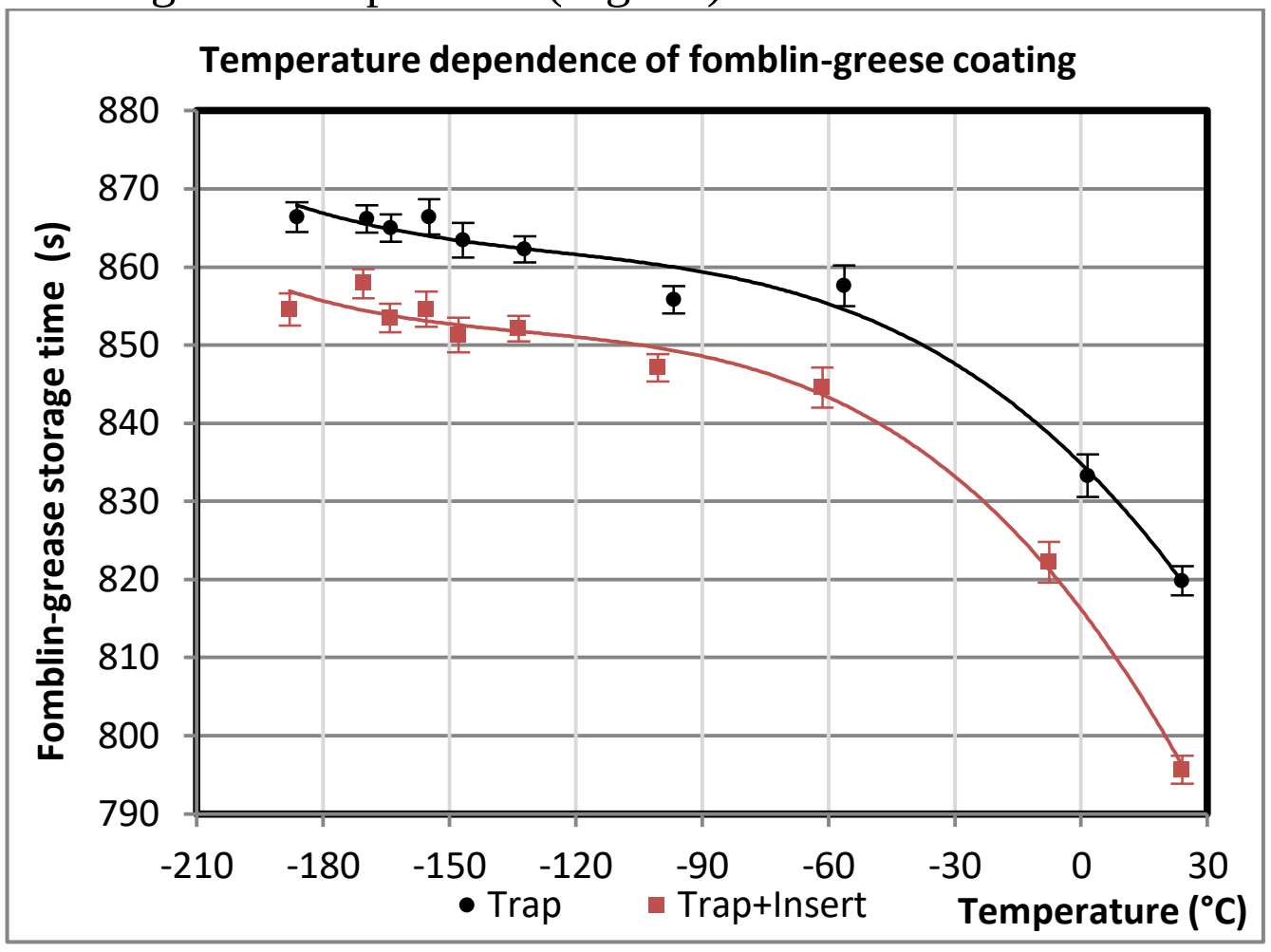

FIG. 11. Temperature dependence of the storage times for the copper trap coated with Fomblin grease.

Another reason to perform measurements at low temperatures is the effect of small heating, where the energy of the neutron is increased during its interaction with the wall of the trap. We observe this effect at room temperature by detecting these upscattered neutrons during the holding period. These are neutrons which have gained enough energy to leave the trap. The small heating effect has been studied with various materials [18] and it was shown that the effect is significantly suppressed at low temperatures. This conclusion is confirmed in our measurements because we do not observe neutron sig-nals above background during the holding period in measurements at about -200°C.

From equation (5) one can obtain the storage time of neutrons under the assumption of zero β-decay probability. Its inverse is equal to the difference between the inverse measured storage time and the inverse free neutron lifetime. This value is an inverse loss probability from all sources, apart from β-decay, and these losses mostly occur at wall collisions. It depends on temperature and this dependence is a characteristic of the quality of the coating (Fig. 12). The dependence thus obtained characterizes the coating quality, in particular the storage time in the trap under the assumption of zero β-decay probability appears to be 60 times longer than the actual free neutron lifetime.

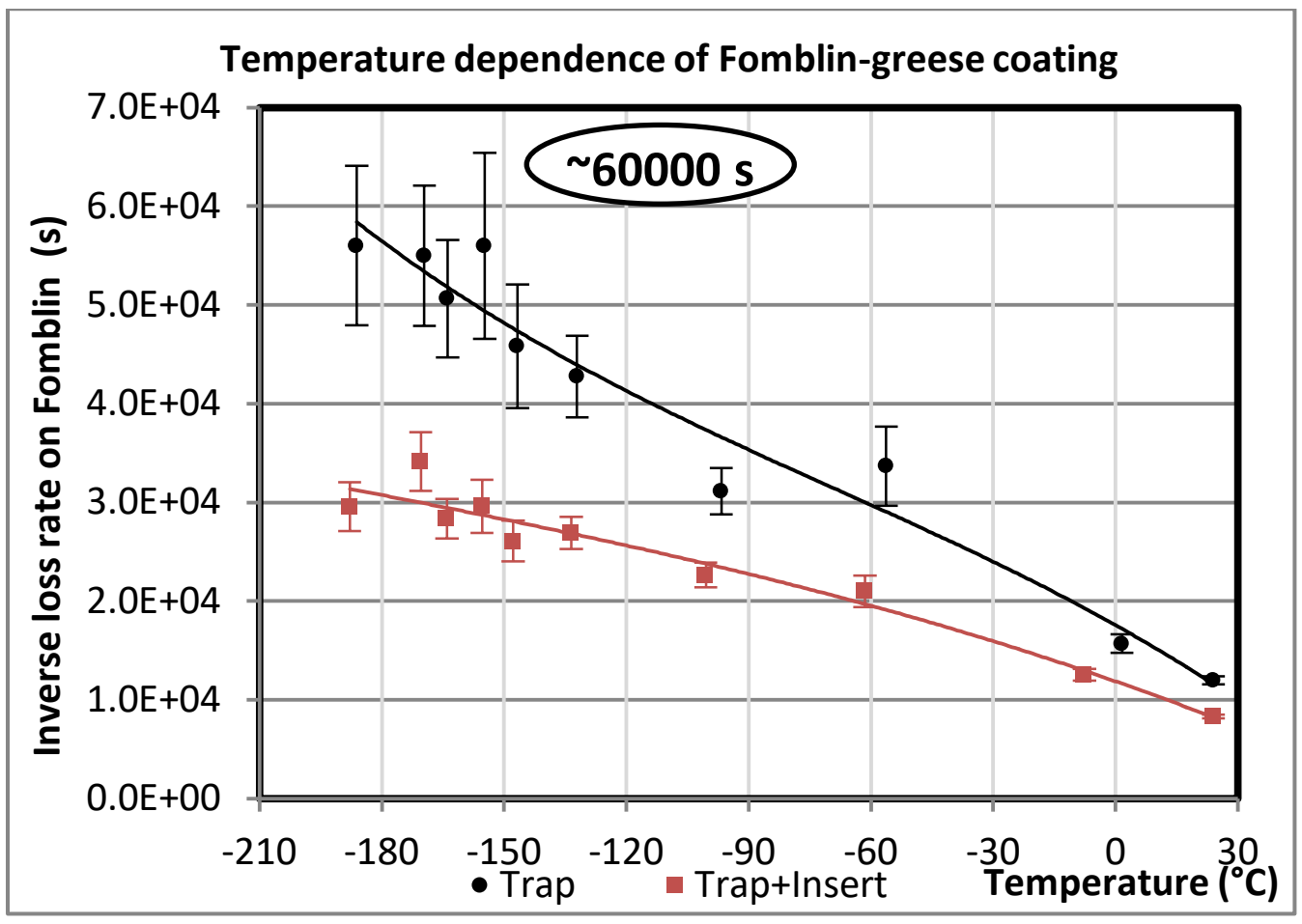

FIG. 12. Inverse UCN loss rate due to the Fomblin grease coating on the copper trap and insert. At liquid nitrogen temperatures, the storage time in the trap for stable neutrons would be 60000 s (~17 hours) so this loss probability is 1.5% of the neutron β-decay probability.

These curves confirm that the coating survives the freezing of Fomblin and further cooling. The loss probability decreases monotonically with temperature. We do not observe any breaks in the curve at the freezing temperature or at lower temperatures and conclude that no macroscopic defects of the covering appear during the cooling.

## 7. Titanium absorber

An essential part of the experimental process is the preparation of the UCN spectrum. Neutrons with energies higher than the gravitational barrier of the trap can bring significant systematic error. At the same time, the total number of registered neutrons determines the statistical accuracy of the experiment and hence it is important to keep more of the useful neutrons in the trap. Until reactor cycle 180, the spectrum was prepared by tilting the trap at 15 degrees for 500 seconds. But this method significantly decreases the number of neutrons in the trap. At 15 degrees many potentially useful neutrons spill over the side and are lost and, in 500 seconds, β-decays reduce the neutron density over the whole spectrum.

In order to speed up the spectrum preparation, we added a titanium absorber to our setup (Fig. 1) at the beginning of cycle 180. It was installed on the insert and rotates with it. Measurements revealed that the absorber reduces the optimal spectrum preparation period from 500 s to 300 s. Also the tilt angle during this phase could be reduced to 10 degrees, enabling the storage of neutrons with energies higher than those that can be stored at a tilt angle of 15 degrees. With this new arrangement, the total number of registered neutrons increased by more than 50%. Count rates for measurements with and without absorber are compared in Fig. 13.

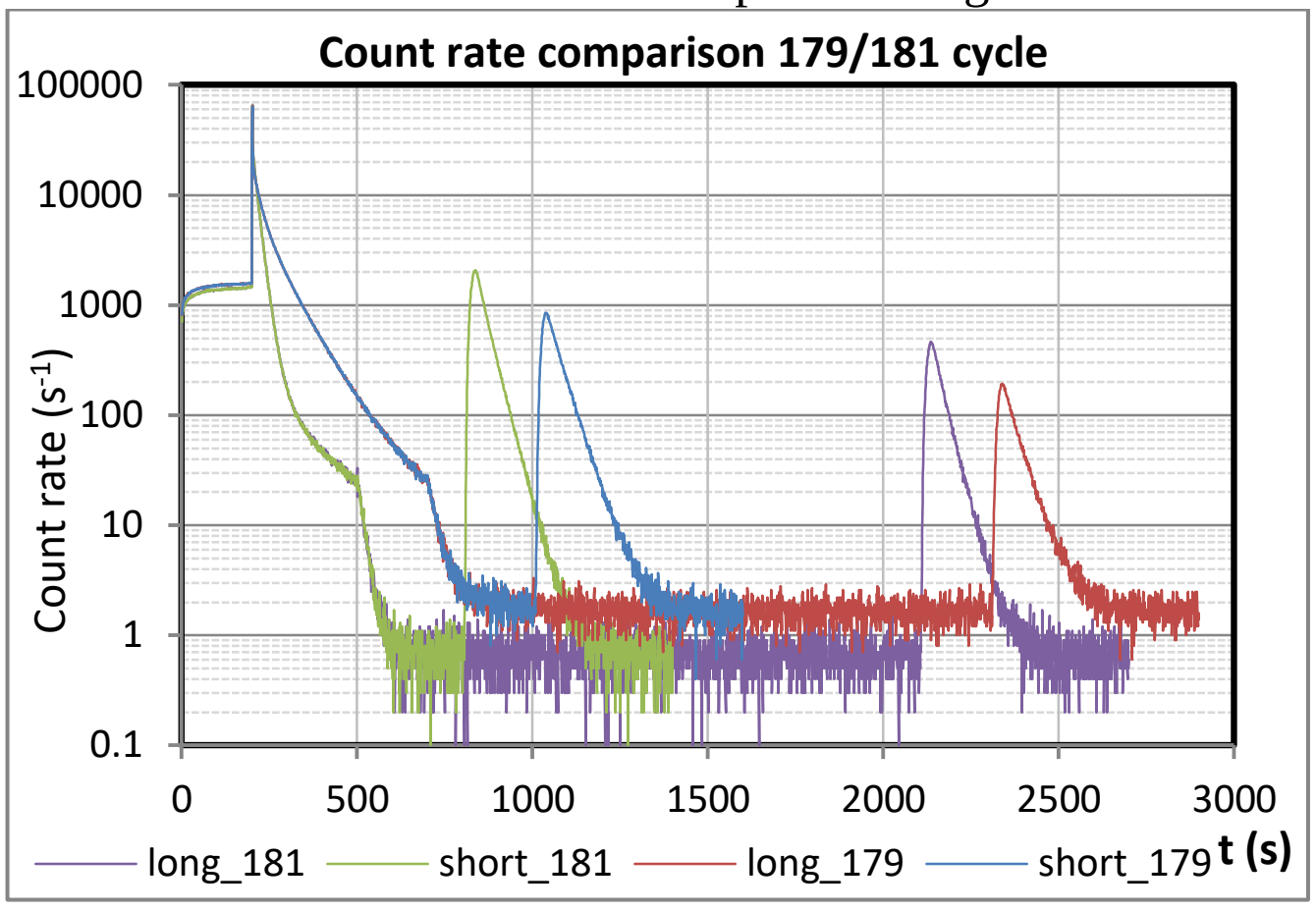

FIG. 13. Neutron count rates before (reactor cycle 179) and after (reactor cycle 181) adding the titanium absorber for swifter spectrum preparation.

## 8. Data analysis

In our experiment the neutron spectrum was divided into two parts by decanting the UCN at two tilt angles. The two geometric configurations therefore give us four points to construct two geometry extrapolations. Further subdivision of the energy spectrum leads to a decrease in the number of detected neutrons in the trap, since each decanting period is 300 or 400 seconds per tilt angle and hence the numbers of neutrons in later decantings decrease due to β-decay.

The principal measurement is the number of neutrons registered by the detector after turning the trap to the tilt angle corresponding to one part of the spectrum. This number depends on the loss rate in the trap and the initial number of neutrons after the filling.

The UCN are detected by a proportional gaseous detector [19, 20]. The detector is multi-sectional and neutrons can be counted independently in each channel. At higher count rates this allows us to decrease the influence of missed counts. However, in order to completely exclude the influence of missed counts, we corrected the neutron count rates using the dead time of the detector. The total error without this correction would be about 0.2 s. In each measurement cycle, the background is measured in the final 200 seconds. The neutron counts are then corrected by subtracting this measured background value. The background depends on the turbine position and actions at other apparatus in the hall, hence we measure it in each measurement and do not use the mean value – we subtract the background of each particular measurement. In the experiment the background/signal ratio is 1% for the short holding process and 5% for the long holding process.

The storage time is obtained from detector counts using equation (6). This method has a significant advantage — the measurements are relative and hence allow us to avoid some systematic uncertainties. For example, we do not need to know the detector efficiency or UCN loss probabilities in the neutron guides because only the ratio of registered neutrons is used. But this is only valid if the initial numbers of neutrons in each measurement cycle are equal. In reality, the initial numbers of UCN are normally distributed, but must be equal within the statistical accuracy required by our experiment. Storage time is obtained without any extrapolation or computer model so its accuracy is mostly statistical and depends on time spent on data taking and the neutron source intensity.

The number of neutrons captured in the filling period slowly decreases during the reactor cycle. This is the result of freezing of thin material films at the junction between the cryogenic part of the apparatus and the neutron guide which is at room temperature. In order to compensate for this "drift" we use a special scheme of measurements. At first we perform the short holding measurement and then the long holding one. On the second set, we swap these so that the long holding is first and the short one is second. For the same reason we alternate measurements with and without the insert.

One can use the mean value as the best estimate of the real physical value of the storage time if the measurements have a normal distribution. To verify the hypothesis of the normal distribution, histograms for each storage time are compared with a normal distribution using the well-known $\chi^2$ criterion. All results are consistent with the hypothesis.

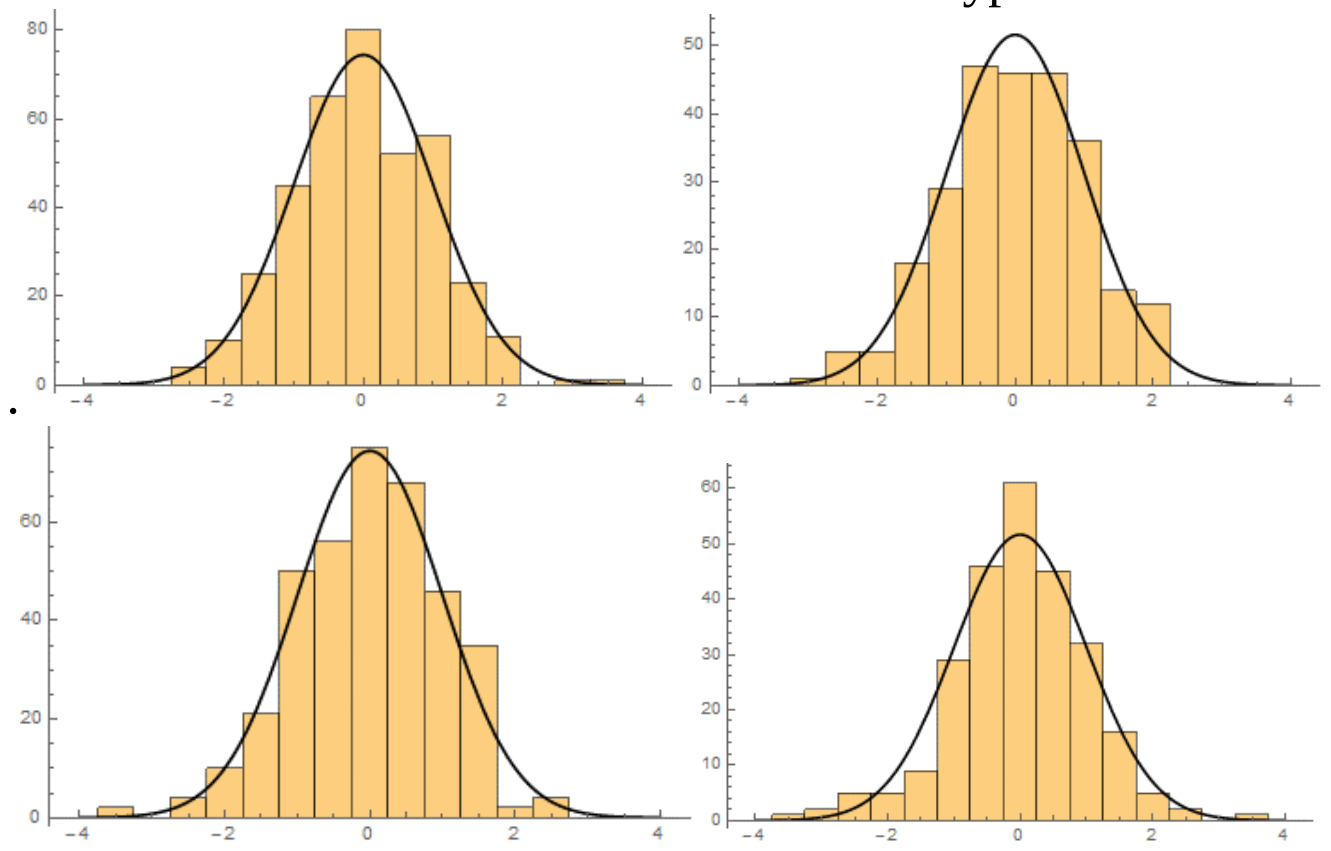

FIG. 14. Normalized histograms of storage times before installation of the titanium absorber. Upper left — first decanting w/o insert ($\chi^2$=1.01), upper right – first decanting with insert ($\chi^2$=0.84), bottom left – second decanting (emptying) w/o insert ($\chi^2$=1.98), bottom – second decanting (emptying) with insert ($\chi^2$=1.53).

The normalized histograms obtained before the installation of the titanium absorber are presented in Fig. 14. Measurements with the absorber are also consistent with a normal distribution.

Having confirmed the normal pattern of measurement distribution we discuss its variance. The distribution width includes two parts: the inherent distribution caused by the probabilistic nature of the decay process; and additional broadening from accidental errors in measurements caused by, for example, reactor power fluctuations, neutron beam fluctuations and drift of UCN beam intensity.

In order to confirm that the main contribution to variance is made by a Poisson distribution of neutron decay, we analytically calculate the distribution width caused by decay only. The uncertainty in measured count rates, using the theory of Poisson distribution, is equal to the square root of the registered number of neutrons. The uncertainty in the storage time was calculated as the error of the value obtained using equation (6). The total uncertainty in the mean value is the square root of the sum of the squared errors divided by the number of considered measurements.

Table 1. Measurements without absorber.

| 178-179 | Value | Measurement uncertainty | Calculated uncertainty |
|---|---|---|---|
| 1 emptying trap only | 862.47 | 0.50 | 0.48 |
| 2 emptying trap only | 865.28 | 0.49 | 0.45 |
| 1 emptying trap and insert | 845.93 | 0.60 | 0.56 |
| 2 emptying trap and insert | 855.33 | 0.59 | 0.54 |

The calculations reveal that the main contribution to the distribution width is from neutron decay. Additional broadening caused by the measurement procedure does not exceed 5%. Storage times and uncertainties are listed in Table 1 and Table 2. The uncertainties are presented to two decimal places in order to emphasize the difference in values.

Table 2. Measurements with absorber.

| 180-181 | Value | Measurement uncertainty | Calculated uncertainty |
|---|---|---|---|
| 1 emptying trap only | 860.33 | 0.55 | 0.51 |
| 2 emptying trap only | 862.80 | 0.56 | 0.53 |
| 1 emptying trap and insert | 842.11 | 0.52 | 0.51 |
| 2 emptying trap and insert | 851.56 | 0.57 | 0.54 |

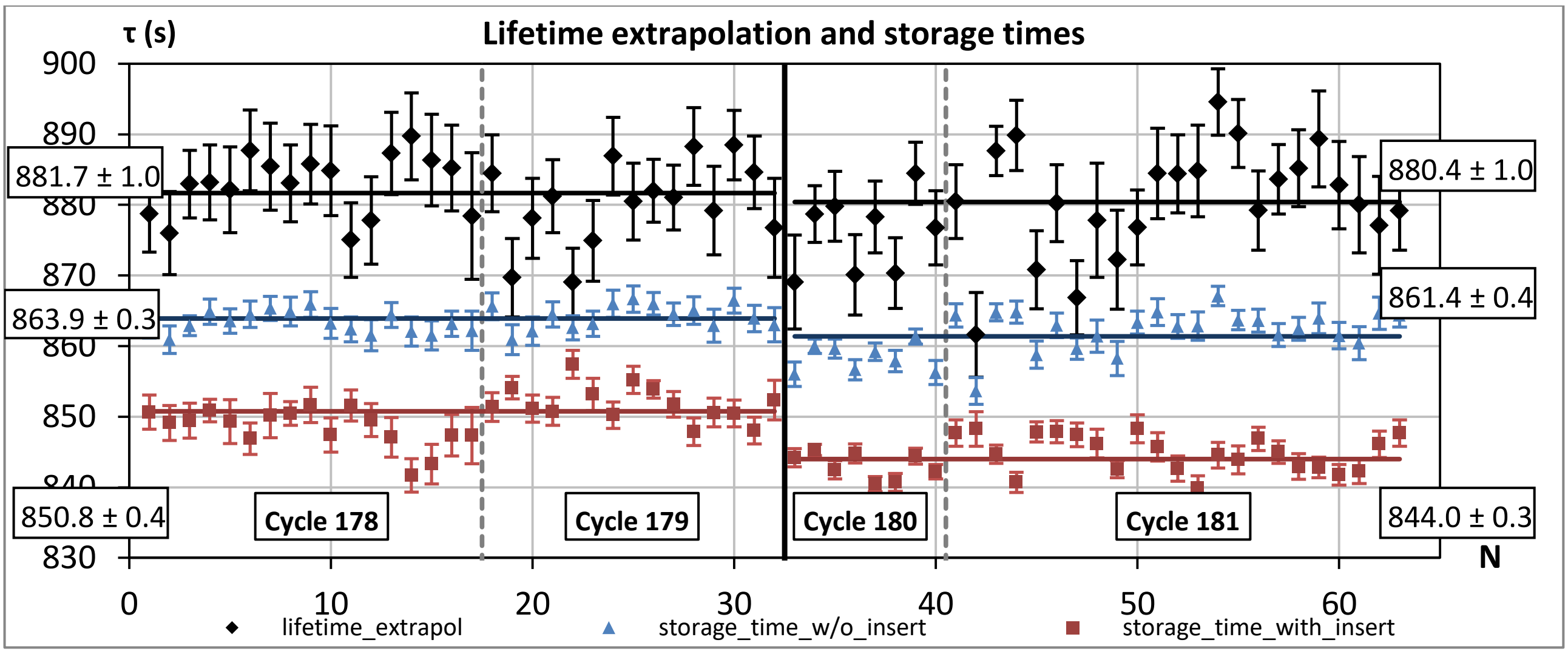

FIG. 15. Time diagram showing successive measurements of the storage times and the corresponding extrapolated free neutron lifetime over the period of the experiment. The vertical black solid line separates measurements with and without the titanium absorber, vertical dashed lines separate the reactor cycles.

Successive measurements (with and without the insert) of the storage times and the corresponding mean lifetime are presented in Fig. 15. Between reactor cycles 179 and 180 the titanium absorber was installed. The mean values are therefore presented separately for measurements with and without the absorber.

After the apparatus was opened, the absorber installed, and then re-closed, the mean storage time in measurements without the insert was reduced by 2.5 seconds and, in measurements with the insert, reduced by 7 seconds. But the neutron lifetime obtained by extrapolation changed only within statistical accuracy, as illustrated in Fig. 15. This is because the free neutron lifetime does not depend on any parameters of our apparatus and its calculation is based on the ratios of storage times and effective collision frequencies.

The significant change of the storage times when moving to measurements with the titanium absorber does not allow us to analyze the combined data. For example the loss coefficient $\eta$ was $(8.44 \pm 0.29) \cdot 10^{-6}$ in measurements without the absorber, and $(9.25 \pm 0.27) \cdot 10^{-6}$ with it installed. Hence, measurements with and without the absorber were treated separately. The final result is averaged over the two regimes. Table 3 lists these results.

Table 3. Extrapolated neutron lifetime.

| Cycle | Lower energies | Higher energies | Averaged |
|---|---|---|---|
| 178+179 | 880.5±1.4 | 882.1±1.3 | 881.4±0.9 |
| 180+181 | 880.1±1.5 | 881.8±1.2 | 881.1±0.9 |
| Averaged | 880.3±1.0 | 881.9±0.9 | 881.3±0.7 |

The extrapolation lines are shown in Fig. 16. The difference between values of the free neutron lifetime obtained by extrapolation for high and low energy UCN does not exceed $1.2\sigma$. Therefore all these measurements of the free neutron lifetime are consistent within statistical accuracy.

In addition to overall extrapolation, we consider the neutron lifetime calculated using data divided into groups. The reason is to observe a possible dynamical time dependence in the measurements. The obtained value of $\chi^2$ allows us to consider the measurements of the neutron lifetime to be normally distributed around a mean value. The mean value obtained by averaging the data obtained through all 4 reactor cycles is presented in Fig. 17. This mean value is consistent with the value obtained by extrapolation within an accuracy of 0.3 s.

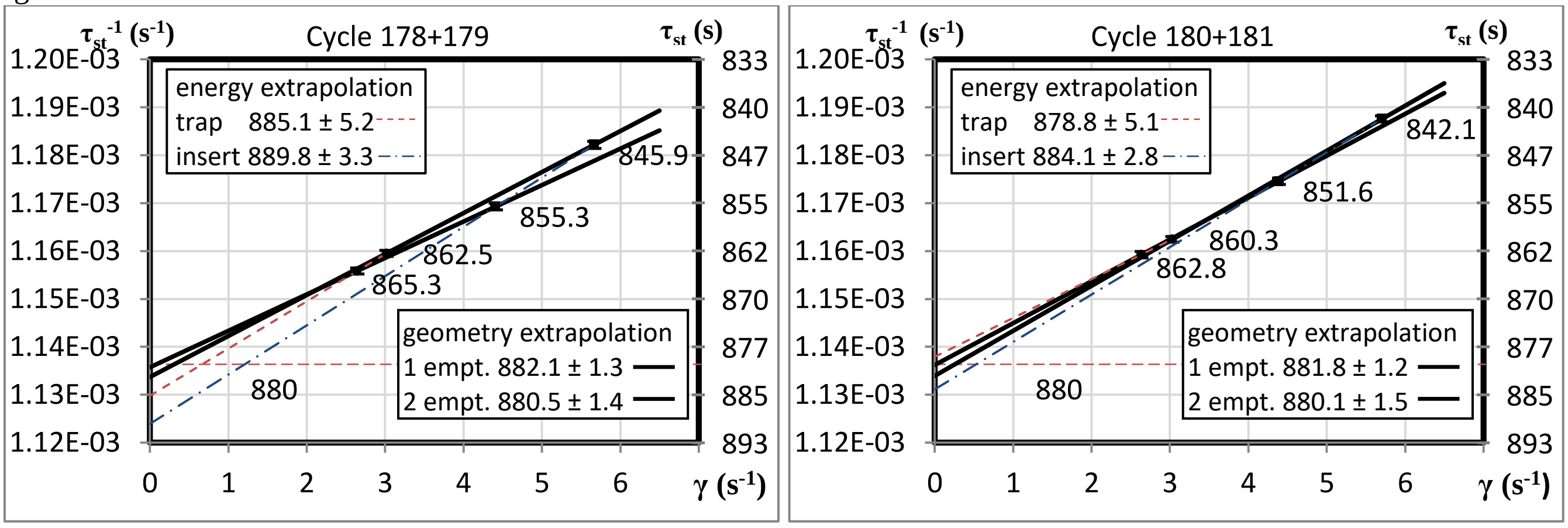

FIG. 16. The extrapolation of measured storage times to the free neutron lifetime. Left — measurements without the titanium absorber, right — measurements with the titanium absorber.

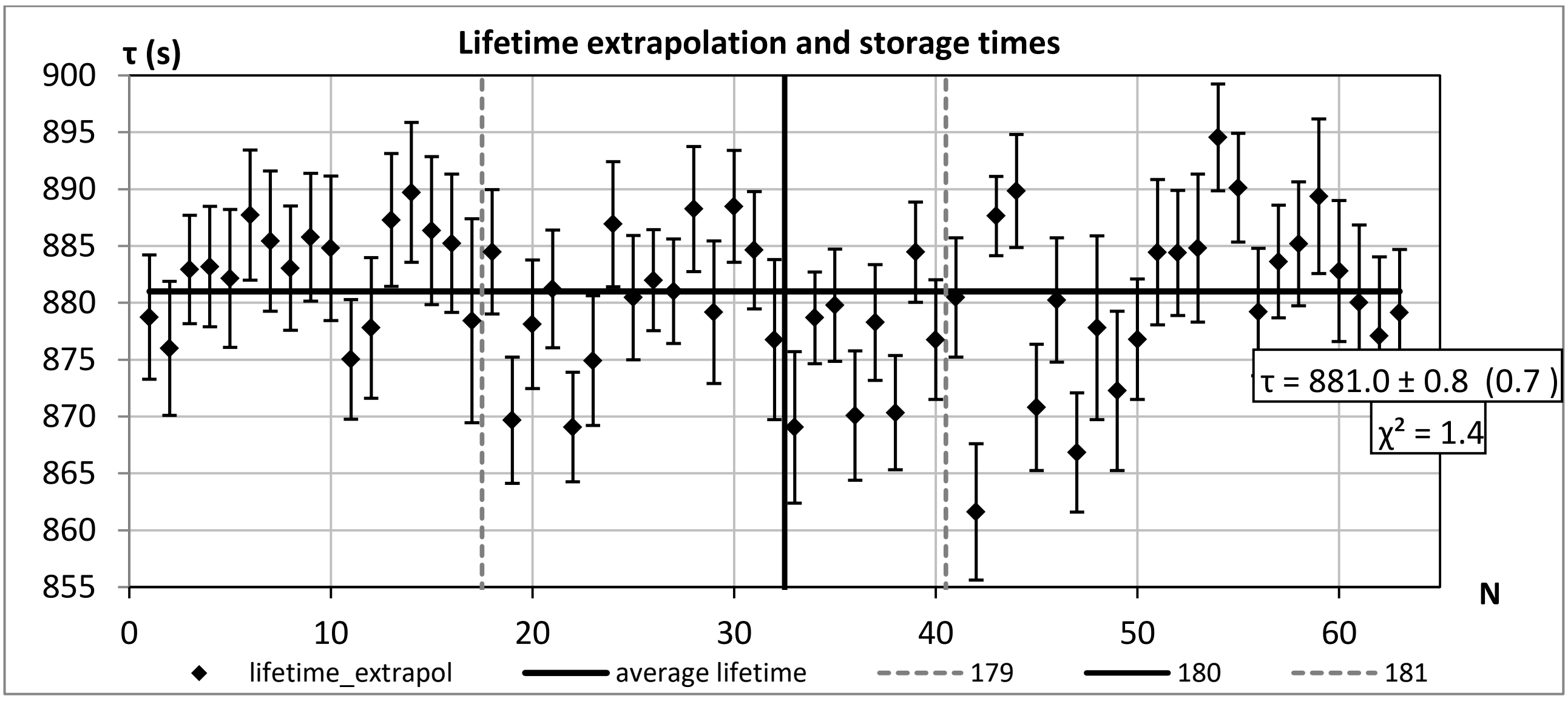

FIG. 17. Time diagram of neutron lifetime measurements.

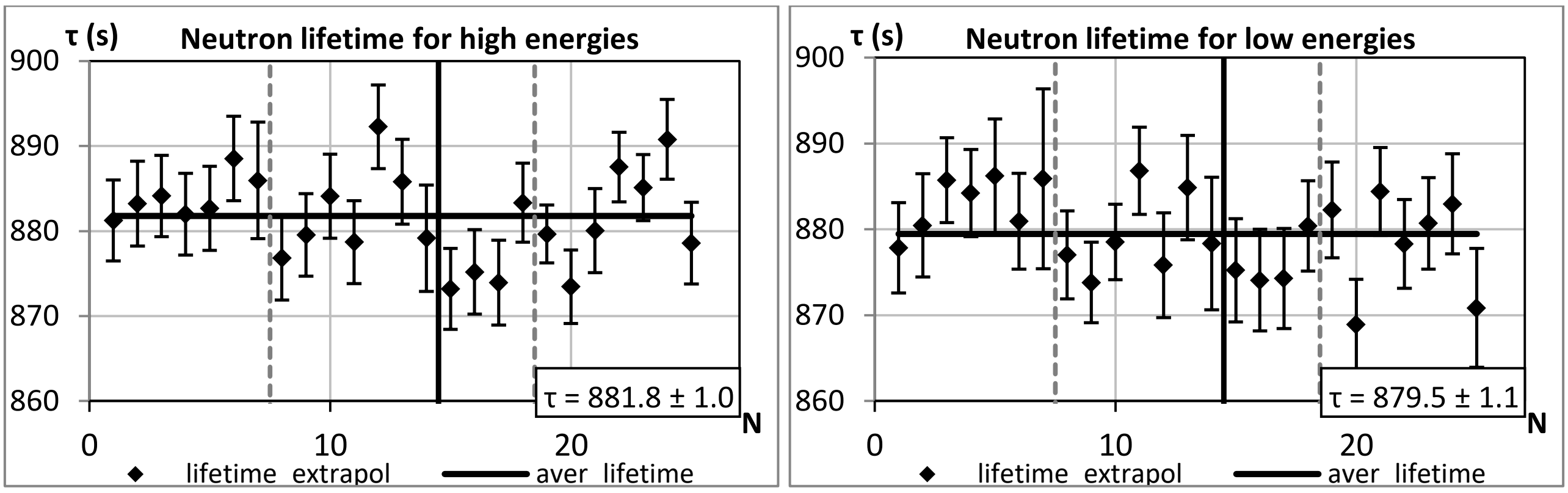

FIG. 18. Time diagram of measurements using single geometry extrapolation. Left — higher energy UCN, right — lower energy UCN.

Time diagrams of neutron lifetimes calculated using geometry extrapolation of either higher energy UCN or lower energy UCN are shown in Fig. 18. Values for the mean lifetime obtained this way have less statistical accuracy, because only half of the data are considered in the calculation. Time diagrams serve to check if there is a time dependence in the measurements and to compare the latest result with the running average. It is important that every storage time and every extrapolation is stable because, although the errors in measuring one storage time have little influence on the averaged value, they may significantly affect the single geometry extrapolation. Self-consistency of all measurements allows us to have confidence in the final result.

The result of measuring the free neutron lifetime with this large gravitational trap of UCN, coated with Fomblin grease, at a temperature of 80 K, and obtained with data gathered in 4 reactor cycles is:

$$\tau_n = 881.3 \pm 0.7\ s$$

This result includes only the statistical accuracy of measurement. In the next section, we analyze systematic uncertainties and corrections to obtain the final result.

## 9. Systematic uncertainties

When calculating the extrapolation to zero collision frequency of UCN with walls of the trap, one must consider sources of possible systematic uncertainties. As we mentioned before, the extrapolation is based on a theoretical model which can only approximate the real process. Most systematic uncertainties originate in assumptions made in section 2.

### a. Uncertainties in $\mu(E)$ function

In our experiment the neutrons interact with a real surface for which the potential $U_0$ can differ from the idealized square barrier. Therefore the actual function $\mu(E)$ will differ from that used in our MC model. In section 4 it was shown that the geometry extrapolation significantly suppresses possible systematic effects caused by the actual function $\mu(E)$ being different to the classical function (8). In calculations using a linear function instead of the classical function (8), the neutron lifetime thus obtained deviates less than 0.2s from that obtained with equation (8). We assume that a deviation 1.5 times higher at a value of 0.3 s can be accepted as the systematic uncertainty caused by obscurity of the true loss function for UCN interactions with the actual surface.

### b. Uncertainty in geometric sizes

The MC model of the neutron path includes geometric parameters of the trap and insert. Errors in these measurements should be taken into account. Geometric parameters affect the simulation and hence the $\gamma$ parameters. We can estimate the errors which originate in the $\gamma$ ratio using expression (14).

$$\Delta\tau_n^{-1} = \sqrt{\frac{x^2(\Delta\tau_1^{-1})^2}{(x-1)^2} + \frac{(\Delta\tau_2^{-1})^2}{(x-1)^2} + \frac{(\tau_2^{-1} - \tau_1^{-1})^2}{(x-1)^4}(\Delta x)^2} \quad (17)$$

Where $x = \gamma_2(E)/\gamma_1(E)$

All distances are estimated to have a maximum uncertainty of 3mm at the operating temperature. If we assume that in the first approximation the ratio of $\gamma_2$ and $\gamma_1$ is proportional to the ratio of surface areas, then $x\sim 2.1$ and $\Delta x \sim 10^{-2}$. In equation (17) the first two terms in the square root correspond to statistical accuracy and the third term describes the contribution of the uncertainty in the γ ratio. This contribution to the systematic error is estimated to be 0.15 seconds.

### c. Uncertainty in the extrapolation method

The MC model allows us to simulate the whole measurement process including the detector count rate. The model requires the free neutron lifetime as an input parameter, and simulated neutron storage times are used for the geometry extrapolation. The difference between the input and the extrapolated value is considered to be a contribution to the systematic error of the geometry extrapolation method. The error obtained by this procedure does not exceed 0.1 s.

### d. Inaccuracy in angle setting

The effective collision frequency is calculated using the MC model which requires parameters such as the geometric configuration, holding periods, and decanting angles. If, in the measurement process, the real angle differs from the one used in the model and hence extrapolation is based on an effective collision frequency which differs from the real one, then it leads to systematic error. To estimate this error an additional calculation was performed. In this calculation the angle of first decanting was altered by ±2 degrees but the extrapolation was made using unaltered $\gamma$ points. The result of the extrapolation is shown in Fig. 19. This size of error in tilt angle cannot lead to a distortion of the extrapolation by more than 0.1 s.

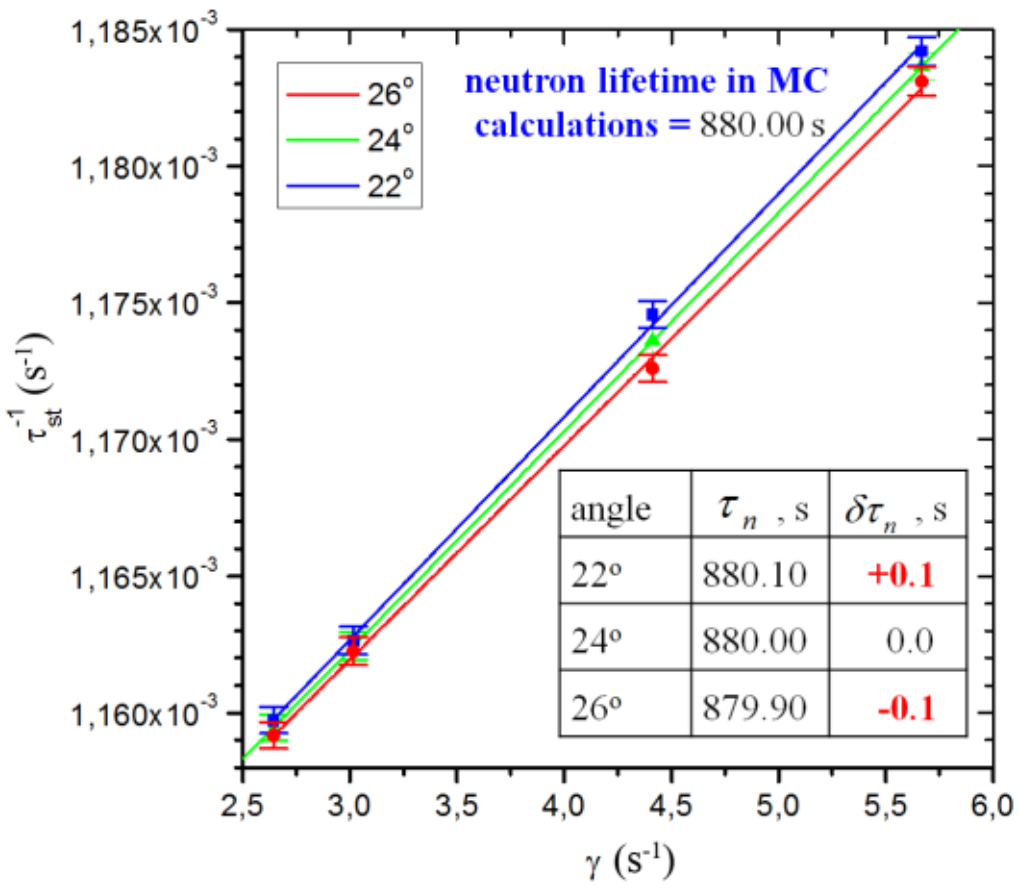

FIG. 19. Uncertainty of trap angular position ±2 degrees.

### e. Inhomogeneity of the covering

The most significant influence on results of extrapolation is due to inhomogeneity in the coating of the trap and insert. In our experiment the trap and insert are copper which has a much smaller probability of UCN capture than titanium.

In section 5 the method to estimate the upper limit of the uncoated area of the surface, which uses data obtained with the titanium trap, was described. The same method of estimating the upper limit is used here. Using the extrapolation method the biggest error appears if all the defects are located on one of the surfaces. Hence, under the assumption that all the defects are located on the insert surface, equations (12) take the form:

$$\begin{cases} \tau_1^{-1} = \tau_n^{-1} + \eta_F \gamma_1 \\ \tau_2^{-1} = \tau_n^{-1} + \eta_F \gamma_2 \dfrac{S_t + S_{ins} - S}{S_t + S_{ins}} + \eta_{cu}\gamma_2 \dfrac{S}{S_t + S_{ins}} \end{cases} \quad (18)$$

where one can see that in the limit of zero defect area S we obtain the original equations (12). In section 5 it was proven that the uncoated area does not exceed $10^{-3}$ of the total surface area of the trap. Assuming that the same value can be applied to the insert, such that $S = kS_{ins}$, the solution of the equations is then:

$$\tau_n^{-1} = \tau_1^{-1} - \frac{\tau_2^{-1} - \tau_1^{-1} - \eta_{cu}\gamma_2 k(1-\alpha)}{\frac{\gamma_2(E)}{\gamma_1(E)} \cdot (1 - k + k\alpha) - 1} \quad (19)$$

Where $\alpha = S_t/(S_t + S_{ins})$ and the difference between $\tau_n$ values obtained with $k = 0$ and $k = 10^{-3}$ is considered as the estimation of systematic error caused by inhomogeneity of the coating. The result of the calculation is 0.5 s.

### f. The influence of the residual gas on UCN storage

In our previous experiment [15, 16] the influence of the residual gas on the measured storage times of the neutrons in the trap was studied. It was shown that a pressure of $3.8 \cdot 10^{-6}\, Torr$ led to a correction of $0.4 \pm 0.02\, s$. The current experiment also holds neutrons in a material trap and the vacuum system is identical to the previous one. The residual gas influence is proportional to the gas density and hence the correction in current measurements is two times smaller since we achieved the pressure of $2 \cdot 10^{-6}\, Torr$ in the trap.

### g. The combined systematic uncertainty

The list of all these effects considered above is presented in Table 4. The result is a systematic error of $0.2 \pm 0.6\, s$.

Table 4. List of systematic effects.

| | Systematic effect | Value, s |
|---|---|---|
| a) | Uncertainty of shape of function $\mu$(E) | ±0.3 |
| b) | Uncertainty of trap dimensions (3 mm for diameter 1400 mm) | ±0.15 |
| c) | Uncertainty of extrapolation method | ±0.1 |
| d) | Uncertainty of trap angular position (2$^{o}$) | ±0.1 |
| e) | Uncertainty of difference for trap and insert coating | ±0.5 |
| f) | The influence of the residual gas | 0.2±0.02 |
| | **Total** | **0.2±0.6** |

## 10. Conclusion

The main advantage of this experiment is the small difference between neutron storage time in the trap and the free neutron lifetime. The large size of the trap and the surface coating which has a small probability of UCN capture allow us to obtain a storage time which is only 15 seconds shorter than the free neutron lifetime. The stability of the coating provides an opportunity to carry out long term experiments with long periods of gathering data. Simulation of the complete neutron paths inside the experimental apparatus allows us to increase the accuracy of the experiment by researching the systematic effects.

After analysis of all the data, we obtain this value for the free neutron lifetime:

$$\tau_n = 881.5 \pm 0.7_{stat} \pm 0.6_{syst}\ s$$

The next step in this experiment is to decrease the temperature of the trap to 10 K to further decrease UCN losses at the walls. This project is being developed at PNPI and reactor time at ILL is scheduled for 2018. The general arrangement of the new apparatus is shown in Fig. 20.

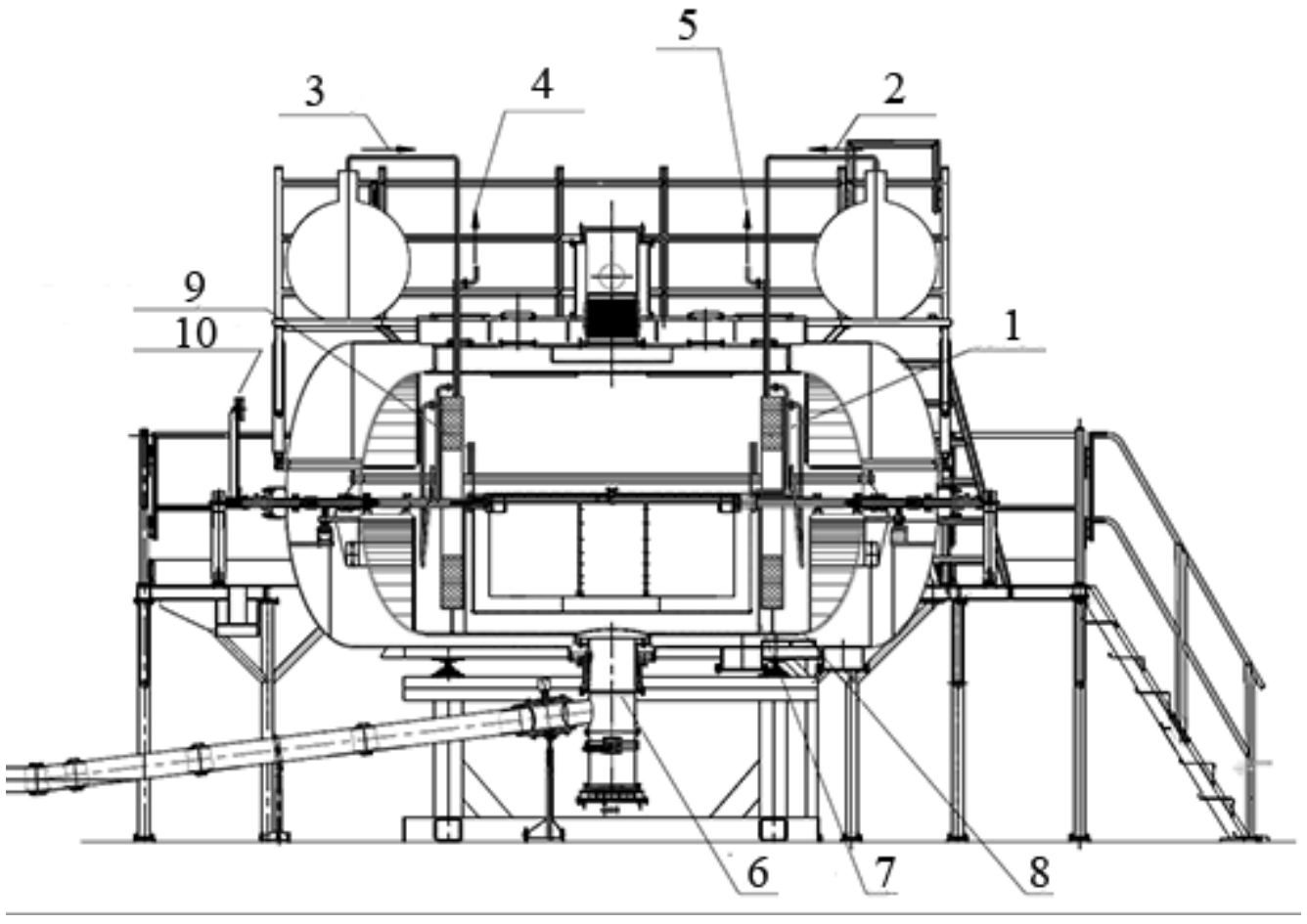

FIG. 20. Conceptual scheme of the new apparatus for the neutron lifetime measurement. 1 — helium tank (volume = 100 liters), 2 — helium trap admission, 3 — Helium insert admission, 4 — helium insert outlet, 5 — helium trap outlet, 6 — aluminum membrane, 7 — barrier, 8 —titanium, 9 — trap counterbalance, 10 — insert counterbalance.

The new apparatus will provide an opportunity to increase the storage time of neutrons by decreasing of the temperature of the trap surface and it will lead to significant decrease of systematic error.

## 11. Comparison of latest results

In the last 20 years several new measurements of the neutron lifetime have been published (Fig. 21). The technology of measuring neutron lifetime using the method of trapped UCN has progressed significantly and provided an opportunity to reach better measurement accuracy. Considering this fact, we would like to present an analysis of results of those most-recent measurements. We considered only results obtained since 2000. Unfortunately, in this period there have been few new results of measuring neutron lifetime using a beam of cold neutrons. Therefore, the discrepancy between the results of storage experiments and beam experiments remains unanswered and requires more data. The value for the neutron lifetime averaged over all these various experiments is $\tau_n = 879.5 \pm 0.8\ s$.

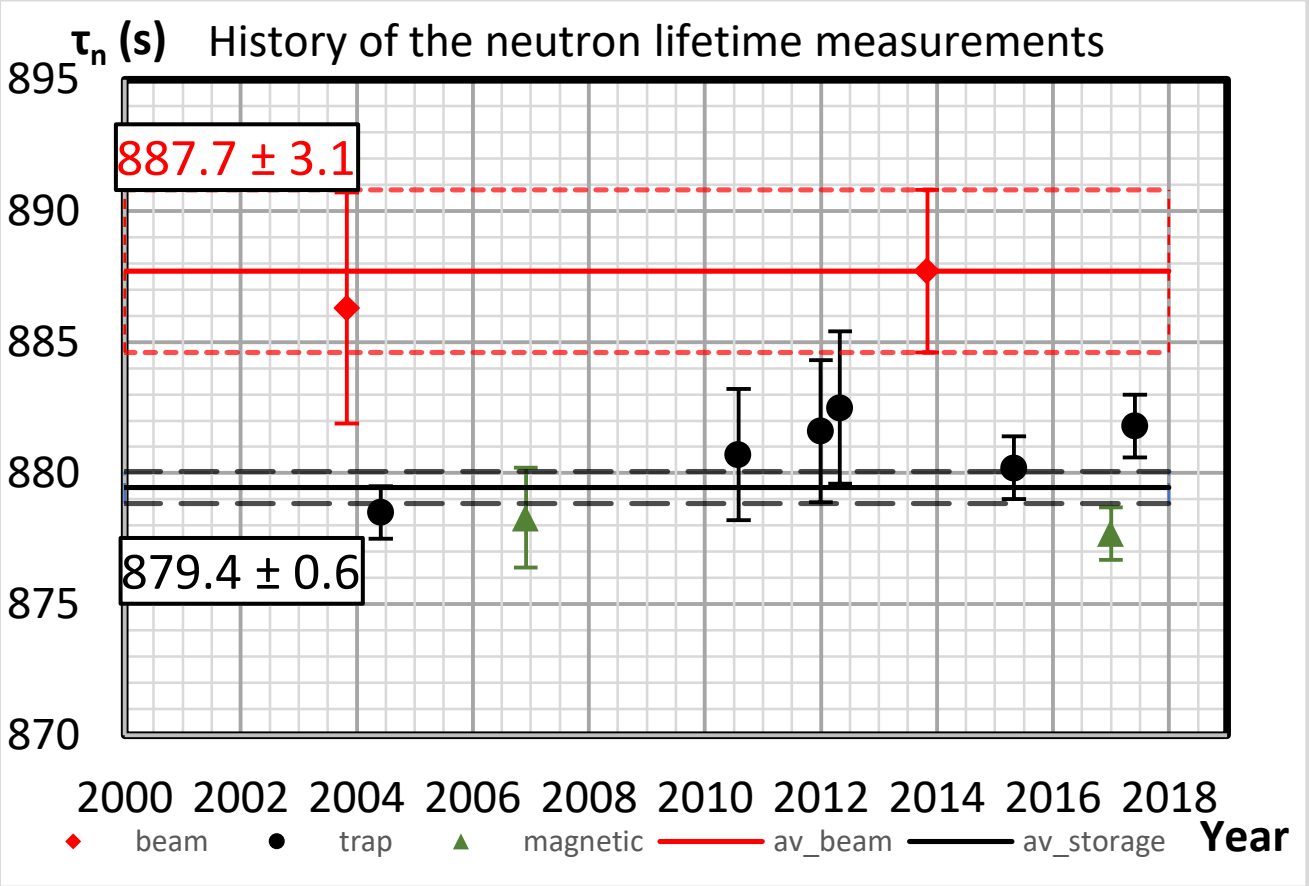

FIG. 21. History of measurements for last 17 years.

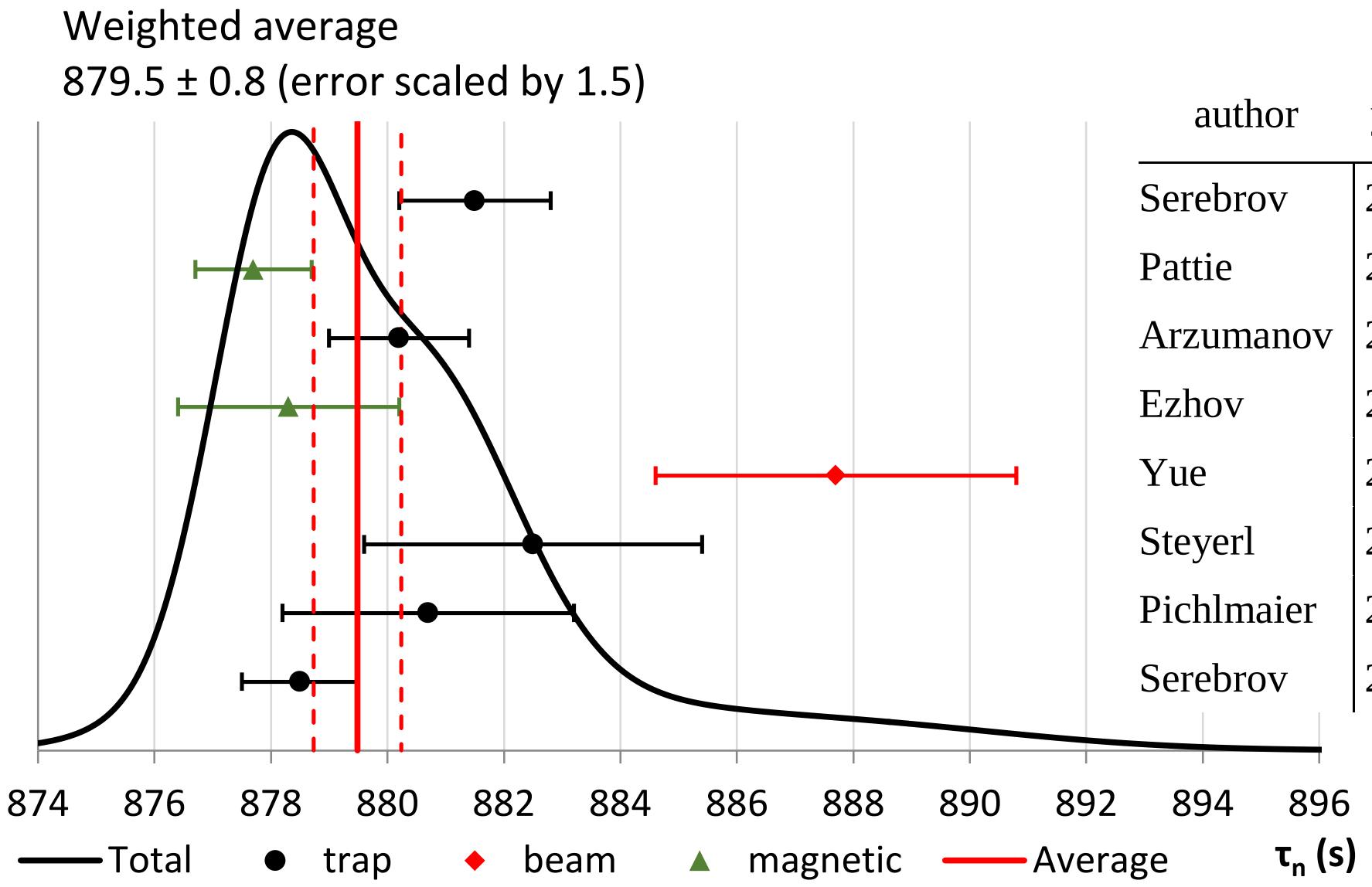

FIG. 22. Distribution of results of measurements for the neutron lifetime.

| author | year | value | error stat | error sys | error Σ | $\chi^2$ | Ref |
|---|---|---|---|---|---|---|---|
| Serebrov | 2017 | 881.5 | 0.7 | 0.6 | 1.3 | 2.4 | — |
| Pattie | 2017 | 877.7 | 0.7 | 0.3 | 1.0 | 3.2 | [21] |
| Arzumanov | 2015 | 880.2 | 1.2 | | 1.2 | 0.4 | [22] |
| Ezhov | 2014 | 878.3 | 1.9 | | 1.9 | 0.4 | [23] |
| Yue | 2013 | 887.7 | 1.2 | 1.9 | 3.1 | 7.0 | [24] |
| Steyerl | 2012 | 882.5 | 1.4 | 1.5 | 2.9 | 1.1 | [25] |
| Pichlmaier | 2010 | 880.7 | 1.3 | 1.2 | 2.5 | 0.2 | [26] |
| Serebrov | 2004 | 878.5 | 0.7 | 0.3 | 1.0 | 1.0 | [15, 16] |

In order to perform more detailed analysis the distribution of experimental results around the mean value was constructed. Using Fig. 22 one can conclude that, although current results of storage experiments have some dispersion, there is no evident contradiction because all the results are in agreement within two standard deviations accuracy if we consider linear summation of systematic errors. The only exception is the beam experiment.

We intend to improve our result by a further continuation of our experiment accompanied by an increase in its accuracy due to adaptation of the installation in order to reach lower trap temperatures of around 10 K. This should reduce the loss factor and allow us to obtain storage times even closer to the free neutron lifetime.

## 12. Acknowledgments

The authors would like to thank the administration of Institute Laue-Langevin for providing neutron facilities for this experiment, and staff members of ILL High-Flux reactor, especially Thomas Brenner, for assistance in installation of the experimental apparatus.

This experiment was supported by the Russian Science Foundation (Project № 14-22-00105)